\documentclass[sigconf]{acmart}

\AtBeginDocument{%
  \providecommand\BibTeX{{%
    \normalfont B\kern-0.5em{\scshape i\kern-0.25em b}\kern-0.8em\TeX}}
    }
\usepackage{wasysym}
\usepackage{url,xspace}
\usepackage{multirow,amsmath}
\usepackage{comment}
\usepackage{xcolor}

\DeclareSymbolFont{letters}     {OML}{cmm} {m}{it}
\usepackage{tikz,pgfplots,pgfplotstable,pgfbaselayers}

\usetikzlibrary{
  shapes.geometric,
  backgrounds,
  positioning,
  arrows,
  external,
  pgfplots.groupplots,
}

\pgfplotsset{compat=1.15}   
\usepgfplotslibrary{
  colorbrewer,
  dateplot,
  fillbetween
}
\pgfplotsset{compat=newest}
\pgfplotsset{
  cycle list/Set1-5,
  cycle multiindex* list={
    mark list*\nextlist
    Set1-5\nextlist
  },
}

\setcopyright{none}
\renewcommand\footnotetextcopyrightpermission[1]{} 
\usepackage{xspace}
\usepackage{xparse}
\usepackage{booktabs}
\usepackage{mathtools}
\usepackage{amsthm}
\usepackage{amsfonts}
\usepackage{scalerel}
\DeclareMathAlphabet{\mathcal}{OMS}{cmsy}{m}{n}
\SetMathAlphabet{\mathcal}{bold}{OMS}{cmsy}{b}{n}
\newcommand{\bnm}{\begin{newmath}}
\newcommand{\enm}{\end{newmath}}

\newcommand{\bea}{\begin{neweqnarrays}}%
\newcommand{\eea}{\end{neweqnarrays}}%

\newcommand{\bne}{\begin{newequation}}
\newcommand{\ene}{\end{newequation}}

\newcommand{\bal}{\begin{newalign}}
\newcommand{\eal}{\end{newalign}}

\newenvironment{newalign}{\begin{align*}%
\setlength{\abovedisplayskip}{4pt}%
\setlength{\belowdisplayskip}{4pt}%
\setlength{\abovedisplayshortskip}{6pt}%
\setlength{\belowdisplayshortskip}{6pt} }{\end{align*}}

\newenvironment{newmath}{\begin{displaymath}%
\setlength{\abovedisplayskip}{4pt}%
\setlength{\belowdisplayskip}{4pt}%
\setlength{\abovedisplayshortskip}{6pt}%
\setlength{\belowdisplayshortskip}{6pt} }{\end{displaymath}}

\newenvironment{neweqnarrays}{\begin{eqnarray*}%
\setlength{\abovedisplayskip}{-1pt}%
\setlength{\belowdisplayskip}{-1pt}%
\setlength{\abovedisplayshortskip}{1pt}%
\setlength{\belowdisplayshortskip}{1pt}%
\setlength{\jot}{-0.4in} }{\end{eqnarray*}}

\newenvironment{newequation}{\begin{equation}%
\setlength{\abovedisplayskip}{4pt}%
\setlength{\belowdisplayskip}{4pt}%
\setlength{\abovedisplayshortskip}{6pt}%
\setlength{\belowdisplayshortskip}{6pt} }{\end{equation}}

\newcounter{ctr}

\newenvironment{newitemize}{%
\begin{list}{\mbox{}\hspace{5pt}$\bullet$\hfill}{\labelwidth=15pt%
\labelsep=4pt \leftmargin=12pt \topsep=3pt%
\setlength{\listparindent}{\saveparindent}%
\setlength{\parsep}{\saveparskip}%
\setlength{\itemsep}{3pt} }}{\end{list}}

\newlength{\saveparindent}
\newlength{\saveparskip}
\DeclareMathSymbol{\mlq}{\mathord}{operators}{``}
\DeclareMathSymbol{\mrq}{\mathord}{operators}{`'}

\newcommand{\E}{{\rm I\kern-.3em E}}

\newcommand{\given}{\ensuremath{\,\big|\,}}

\newcommand{\getsr}{{\:{\leftarrow{\hspace*{-3pt}\raisebox{.75pt}{$\scriptscriptstyle\$$}}}\:}}
\newcommand{\genfrom}[1]{{\:{\leftarrow{\hspace*{-3pt}\raisebox{.75pt}{$\scriptscriptstyle#1$}}}\:}}

\newcommand{\secref}[1]{\mbox{Section~\ref{#1}}}

\newcommand{\figref}[1]{\mbox{Figure~\ref{#1}}}

\renewcommand{\eqref}[1]{\mbox{(\ref{#1})}}

\newcommand{\gamesfontsize}{\footnotesize}
\newcommand{\funcfont}[1]{{\textsf{#1}}}

\newcommand{\fpage}[2]{\framebox{\begin{minipage}[t]{#1\textwidth}\gamesfontsize  #2 \end{minipage}}}

\def \part {part}

\newcommand{\pw}{w}

\renewcommand{\paragraph}[1]{\vspace*{6pt}\noindent\textbf{#1}\;}

\newcounter{mytable}
\def\mytable{\begin{centering}\refstepcounter{mytable}}
\def\endmytable{\end{centering}}

\newcounter{myfig}
\def\myfig{\begin{centering}\refstepcounter{myfig}}
\def\endmyfig{\end{centering}}

\def \blackslug{\hbox{\hskip 1pt \vrule width 4pt height 8pt
    depth 1.5pt \hskip 1pt}}
\def \qed{\quad\blackslug\lower 8.5pt\null\par}

\newcounter{mynote}[section]

\newcommand\ignore[1]{}

\newcounter{rcnote}[section]

\newcounter{mmnote}[section]

\newcommand{\mytab}{\hspace*{.4cm}}

\newcommand{\rhf}[2]{R_{f, \gamma}}

\newcommand{\w}{{w}}
\newcommand{\pp}{s}

\DeclareDocumentCommand{\edist}{o o}{
  \ensuremath{
    \IfNoValueTF{#1}{{d}}{{\sf d}(#1,#2)}
  }
}

\newcommand{\mathcmd}[1]{\ensuremath{#1}\xspace} %

\newcommand{\PW}{\mathcmd{W}}

\DeclareMathSymbol{\mlq}{\mathord}{operators}{``}
\DeclareMathSymbol{\mrq}{\mathord}{operators}{`'}

\newcommand{\ours}{\funcfont{Mascara}\xspace}
\newcommand{\markov}{\funcfont{Markov}\xspace}
\newcommand{\wiki}{\funcfont{Wiki-5}\xspace}
\newcommand{\pwlen}{l}

\usepackage{ifthen}

\newcommand{\MP}[1]{\textcolor{red}{#1}}

\newboolean{publicversion}
\setboolean{publicversion}{false}

\ifthenelse{\boolean{publicversion}}{
   \newcommand{\grumbler}[2]{}
   
}{
   \newcommand{\grumbler}[2]{\MP{{\bf #1: #2}}}

}

\newcommand{\oldText}[1]{}

\newcommand{\system}{\textrm{\sf {MASCARA}}\xspace}

\newcommand{\cer}{\text{CER}}

\newcommand{\sdchr}{\sigma_\text{chr}}
\newcommand{\lprob}{L_{1}}
\newcommand{\lprobbi}{L_{2}}
\newcommand{\model}{\mathcal{M}}
\newcommand{\start}{\ensuremath{\mathsf{<\text{s}>}}}
\newcommand{\lastword}{\ensuremath{\mathsf{<e>}}}
\newcommand{\commentNew}[1]{{\color{gray}{\scriptsize\ /* #1 */}}}
\newcommand{\sysgen}{\ensuremath{\mathsf{MascaraGen}}}
\newcommand{\mkmodel}{\mathcal{M}}
\newcommand{\sw}{B}
\newcommand{\score}{S}
\newcommand{\thetaone}{\theta_1}
\newcommand{\thetatwo}{\theta_2}
\newcommand{\nextstr}{\mathsf{next}}

\newcommand{\nppws}{72,999}

\newcommand{\dice}{\funcfont{Diceware}\xspace}
\newcommand{\userpp}{\funcfont{User}\xspace}
\newcommand{\mmap}{\funcfont{TemplateDice}\xspace}

\newcommand{\changed}[1]{{\color{black} #1}}

\makeatletter
\renewcommand{\paragraph}{%
  \@startsection{paragraph}{4}%
  {\z@}{0.5ex \@plus 1ex \@minus .2ex}{-1em}%
  {\normalfont\normalsize\bfseries}%
}
\makeatother

\fontencoding{T1}
\begin{document}

\date{}

\title[\system{}: Generating memorable and secure passphrases]{\system{}: Systematically Generating \\Memorable And Secure Passphrases}

\DeclarePairedDelimiter{\ceil}{\lceil}{\rceil}

\author{Avirup Mukherjee }
\affiliation{%
  \institution{IIT Kharagpur}
  \country{}
}
\email{avimukh.250696@iitkgp.ac.in}

\author{Kousshik Murali}
\affiliation{%
  \institution{IIT Kharagpur}
  \country{}
}
\email{kousshikraj.raj@gmail.com}

\author{Shivam Kumar Jha}
\affiliation{%
  \institution{IIT Kharagpur}
   \country{}
}
\email{shivam.cs.iit.kgp@gmail.com}

\author{Niloy Ganguly}
\affiliation{%
  \institution{IIT Kharagpur, L3S Hannover}
   \country{}
}
\email{niloy@cse.iitkgp.ac.in}

\author{Rahul Chatterjee}
\affiliation{%
  \institution{University of Wisconsin--Madison}
   \country{}
  }
  \email{rahul.chatterjee@wisc.edu}

\author{Mainack Mondal}
\affiliation{%
  \institution{IIT Kharagpur}
   \country{}
}
\email{mainack@cse.iitkgp.ac.in}

\renewcommand{\shortauthors}{}

\begin{abstract}

Passwords are the most common mechanism for authenticating users online. However, studies have shown that users find it difficult to create and manage secure passwords. To that end, passphrases are often recommended as a usable alternative to passwords, which would potentially be easy to remember and hard to guess. However, as we show, user-chosen passphrases fall short of being secure, %
while state-of-the-art machine-generated passphrases are difficult to remember.

In this work, we aim to tackle the drawbacks of the systems that generate passphrases for practical use. In particular, we address the problem of generating secure and memorable passphrases and compare them against user chosen passphrases in use. We identify and characterize $\nppws$ user-chosen in-use unique English passphrases from prior leaked password databases. Then we leverage this understanding to create a novel framework for measuring memorability and guessability of passphrases. Utilizing our framework, we design \system{}, which follows a constrained Markov generation process to create passphrases that optimize for both memorability and guessability. Our evaluation of passphrases shows that \system{}-generated passphrases are harder to guess than in-use user-generated passphrases, while 
being easier to remember compared to state-of-the-art machine-generated passphrases.
We conduct a two-part user study with \changed{crowdsourcing platform Prolific} to demonstrate that users have highest memory-recall (and lowest error rate) while using \system{} passphrases. Moreover, for passphrases of length desired by the users,
the recall rate is 60-100\% higher for \system-generated passphrases compared to current system-generated ones. 
    
\end{abstract}



\maketitle
\pagestyle{plain}

\urlstyle{tt}
\section{Introduction}\label{sec:intro}
Passwords are by far the most popular method for authentication, despite their several limitations: %
users have to remember many (unique) passwords and, in turn, they tend to choose weak, easy-to-guess passwords. 
\changed{To improve security of user passwords, several alternative tools and strategies are proposed, such as password managers and two-factor authentication mechanisms.  %
However, to secure password managers (and even while two-factor authentication is used), users are required to create and remember a secure and memorable master secret.
To that end, \emph{passphrases} are considered an alternative approach to generate memorable yet strong authentication secrets. }

\changed{Passphrases, unlike passwords, are sequences of words, for example ``correct horse battery staple''. %
They are frequently recommended for particularly sensitive scenarios---as the master secret for password managers~\cite{huth2012password}, for locking users' Cryptocurrency wallets (e.g., brainwallet~\cite{eskandari2018first}), or protecting SSH keys~\cite{ssh-keygen}.
Passphrases could also be used to create more memorable passwords. For example, mnemonics created from
passphrases can assist in memorizing complex passwords~\cite{woo2016improving}.
Thus Passphrases are a secure and user-friendly alternative to passwords~\cite{fbi-2020-passphrase}, while we (slowly) transition toward a ``password-less'' future of authentication.}

Passphrases can be user-generated 
or system-generated. User-generated passphrases are memorable, often because of their sentence/grammatical structure being
aligned with natural text~\cite{gram1, gram4} %
To that end, system-generated passphrases are proposed---such passphrases are difficult for an attacker to guess~\cite{bonneau2012quest}, but are also difficult to remember~\cite{jagadeesh2021alice, shay2012correct}.

Existing approaches to computer-generated passphrases include (a) \emph{Diceware}~\cite{reinholddiceware}, which picks random words from a given wordlist, and (b)  \textit{Template-based Diceware}~\cite{mmapCode}, which generates passphrases that adhere to a small set of pre-selected syntax template (e.g., user-adverb-verb-noun).  Shay et al. \cite{shay2012correct} showed that Diceware passphrases are 
as difficult to remember as randomly generated passwords. 
Further, our analysis reveals (\secref{ssec:systempp}) that Template-based Diceware also suffers two major drawbacks: (1) there is a fixed maximum bound on the strength of the generated passphrases, because it uses a handful of specific syntax templates, and (2) there is no systematic way to extend the space of passphrases with more templates (\secref{ssec:langmodel}), necessitating a thorough re-design of the approach. Another prior work took an alternative approach---they tried to use techniques like implicit learning \textit{post-passphrase generation} to improve memorability~\cite{jagadeesh2021alice}.

Complementary to those efforts of enhancing memorability by user-learning, we ask: \textit{Is it possible to develop a simple automated approach for producing system-generated passphrases of arbitrary length, which is memorable by abiding grammar/sentence structure, yet hard for an adversary to guess and address shortcomings of existing passphrase-generation systems?}

In this work, we answer this question affirmatively and present \system{}, which can automatically generate memorable passphrases with good security. \changed{We note that the current passphrase generation methods like Diceware implicitly trade off memorability for security. In this work, we explore this trade-off and design, build, and evaluate \system{} which attempts to provide a balanced trade-off between security and memorability.} We design \system{} using insights from a novel in-use English passphrase dataset to ensure good guessability-memorability trade-offs. 
The key contributions of this work are:
\begin{newitemize}
    \item We used heuristics to identify $\nppws$ user-chosen English passphrases %
    from prior password leaks. To the best of our knowledge, our dataset is the largest user-chosen passphrase dataset to date.\footnote{The dataset and code for this work can be found at \color{blue}{\tt\url{https://github.com/Mainack/MASCARA-passphrase-code-data}}.} Our algorithm leverages word segmentation in the noisy text to identify these passphrases. The syntactic structures of these passphrases are distinctly different from \dice---favoring memorability over guessability. %
    
    \item We created a memorability-guessability measurement framework for passphrases. Building on prior works on the memorability of natural language phrases we identified distinct and important features of a sentence %
    which affects the memorability of a passphrase. We additionally created a Monte-Carlo estimate of the passphrase guess ranks to measure the guessability of passphrases. %
    We utilize this framework to balance memorability and guessability during passphrase generation.

    \item Using our framework, we present \system{}, a novel system to automatically generate memorable yet not so easily guessable passphrases. \system{} leverages a constrained generative process by modifying a generative Markov model and explicitly considering the dimensions of memorability and guessability during passphrase generation. Our evaluation demonstrated that \system{} generated passphrases have improved security than user-generated ones and do not suffer from any of the drawbacks of the existing systems. Moreover, the user study we carried out shows that users have the highest recall rate after two days for passphrases in their preferred range with \system{}. %
    For passphrases of length 7 or less (preferred by most users) \system{} provides 1.6x--2x better recall rate than deployed systems like \dice while maintaining a less than 10\% character error rate. 
\end{newitemize}

\noindent \textbf{Limitations:} Our study has three key limitations. First, we are not providing a one-step solution to the quest of memorable passphrases. Rather \system{} takes a principled and complementary approach to enhance today's system-generated passphrases by balancing memorability and security. %
\changed{Second, as with any user study, ecological validity is hard to ensure objectively. In line with earlier work, we made a conscious and earnest effort in our experiments to not nudge our participants to choose or better remember passphrases from any specific system (to preserve the sanctity of recall values)~\cite{egelman2013ecologicalpass}. Participants did not know passphrases shown to them came from which system---our quantitative and qualitative analysis also does not indicate any bias. %
In fact, if the users believed that our study required to remember their chosen passphrases correctly and thus always choose the one which was easiest to remember and go out of their way to memorize (e.g., using pen and paper), we might not have seen the wide variance in recall values (see~\figref{fig:recall_cer_less7}).}
Finally, in line with prior work, we only consider English passphrases---it enables comparison with state-of-the-art~\cite{jagadeesh2021alice,
bonneau_linguistic_2012, bonneau_2018}. \changed{Exploring passphrases for other languages is part of our future work.}

\section{Background and related work}
\label{sec:relwork}

\subsection{System-generated Passphrases}

\noindent Passwords are used for authenticating to critical as well as non-critical infrastructures~\cite{bonneau2012quest}. Unfortunately, several prior works~\cite{das2014tangled} 
have shown  that users tend to choose predictable passwords and reuse them across multiple accounts. 
To that end, in the past decade, passphrases have been put forward as an alternative or complementary mechanism to passwords. NIST defined a passphrase to be a \textit{memorized secret consisting of a sequence of words or other text that a claimant uses to authenticate their identity}~\cite{grassi2017draft}. Intuitively, passphrases are likely to be easier to remember than passwords (due to their closeness to natural language) for users as well as harder to guess (due to their length) for adversaries. Shay et al.~\cite{shay2012correct} demonstrated (using a user study) that even simply using passphrases consisting of three to four words can be comparable in terms of entropy with passwords generated using more-involved methods while also accounting for memorability, highlighting the utility of good passphrases. 
Today passphrases are used in password managers, cryptocurrency wallets~\cite{8285737}, and for securing ssh keys~\cite{ssh-keygen}. 

\changed{Several prior works focused on generating and remembering secure passwords (and passphrases) using techniques like contextual cues, portmanteau, or mnemonic based generation~\cite{stobert2014password, woo2020we,kuo2006human, al2015towards,joudaki2018reinforcing}. These techniques aim to associate a context (like an image) with the secret. Consequently, our work
on generating secure and memorable passphrases is complementary to such techniques of remembering secrets---they can be directly used to improve memorability of \system-generated passphrases.}

Prior works however have not systematically investigated the guessability-memorability trade-off. %
 In this work, we formalize the guessability-memorability problem  by creating a data-driven framework and leverage the framework to design  \system{} that can generate secure and memorable passphrases.

\subsection{Security and threat model}
\label{ssec:secandthreat}

\noindent Measuring security of passwords is a well-studied topic~\cite{klein1990foiling}. Earlier works considered different approaches the attacker could employ and thus used Markov models, probabilistic CFG, neural networks, etc. for the guessability estimations~\cite{wang2016targeted, pal2019beyond,
van_acker_password_2015, melicher-meter, ur-measure}. However, there is relatively less work in the domain of passphrases. 

\vspace{4mm}

\noindent \textbf{Using guess rank to measure passphrase security} Previous works found that the security of passphrases can be increased by increasing the entropy via either introducing semantic noises or increasing the wordlist size~\cite{lee2007passphrase}. However, if users are given control, they generally tend to opt for common phrases, reducing security drastically~\cite{kuo2006human}. In all of these works, the security of passphrases is generally measured through either entropy or user surveys~\cite{renyi1961measures}. 

But later, researchers have shown that \textit{guess rank} is a much more acceptable measure than entropy for measuring the security of a password~\cite{realWorldAccuracies, testingMetrics}. The guess rank of a password can be understood as the number of guesses an adversary needs to arrive at the correct password. So higher the guess rank, lower the guessability. Building upon this and earlier works on password meters~\cite{van_acker_password_2015, melicher-meter, ur-measure}, we use the guess rank metric for measuring the guessability of passphrases in our setup. Specifically, we estimate a guess rank for each passphrase in our setup--- the higher the guess rank, the higher is its resistance to external attacks. 

\vspace{4mm}

\noindent \textbf{Our adversary model with \textit{min auto} approach.} \changed{In this work, we consider a powerful offline generalized untargeted adversary.  We assume the attacker is fully aware of the passphrase generation algorithm and the dataset used in all training. The attacker can generate (offline) as many passphrases as they can (given their computational resources) for any algorithm.
The goal of the attacker is to guess the password randomly generated using an algorithm given a large guessing budget (e.g., $10^{15}$). Our offline attack~\cite{bonneau2012quest} model is stronger than an online attack setting where the attacker is limited by the number of guesses they can make. }
The primary challenge for the attacker is to generate an ordered list of passphrase guesses $\pw_{1}, \pw_{2}, ..., \pw_{n}$ to reach the target passphrase $\pw$ as early as possible (least number of guesses). The higher the guess rank of a passphrase the stronger the passphrase is (lower guessability). Given there are multiple algorithms an attacker can use to guess a passphrase we take a \textit{min auto} approach described by Ur et al.~\cite{realWorldAccuracies}. First, we used multiple password cracking algorithms ($n$-gram word and character models trained over a large corpus of passphrases generated by the system under consideration),  parameterized by a set of training data~\cite{Kelley2012, Bonneau2012}. For each algorithm, the \emph{guess rank} will be the number of incorrect guesses the particular algorithm used to arrive at the correct passphrase. Since the average guess rank for even a passphrase of medium strength is of the order of $10^{15}$ (\secref{ssec:guesseval}), running the algorithms to find the guess rank is infeasible---we therefore used Monte Carlo simulations to estimate the guess rank (\secref{ssec:guessability_old}). Finally, following the previous work by Ur et al. we simply took the minimum of all guess ranks to arrive at our estimated guess rank. Ur et al. demonstrated that taking minimum guess rank of all automated approaches (called \textit{min auto}) is a reasonable approximation of real-world cracking scenario~\cite{realWorldAccuracies}.

\subsection{Measuring memorability of passphrases}
\label{ssec:memorablemeasure}

\noindent Earlier works tried to correlate password memorability with the frequency of passwords, login durations, and even keyboard pattern~\cite{gao2018forgetting, guo2019optiwords}. Other works used methods like encoding random n-bit strings for generating memorable passwords or using chunks~\cite{ghazvininejad2015memorize}. %
All of these cases measured memorability of passwords based on user surveys instead of an automated linguistic metric~\cite{chatterjee2017typtop, woo2020we, shay2012correct}. Although useful, these works considering memorability of passwords are complementary from that of passphrases. Passphrases often contain possible linguistic properties (the order in which words are presented) which are generally absent in passwords. To that end, there is some work on the memorability of English phrases. For example, work by Danescu et. al.~\cite{danescu2012you} has tried to measure memorability of popular movie quotes using lexical distinctiveness. Some Human-Computer Interaction studies identified Character Error Rate (CER) as an acceptable measure for memorability of passphrases~\cite{mackenzie2002character, kristensson2012performance}. We build on this research, identify underlying factors affecting memorability of phrases, and consequently optimize to improve memorability of  \system{}-generated passphrases (\secref{ssec:memorability} and \secref{ssec:tuning}). Our two-part user study results demonstrated that users indeed memorize \system{} generated passphrases which leverages this model. With this background, in the next section, we start with first identifying and analyzing the state of the art methods for generating passphrases in the wild. %

\section{Understanding Passphrase generation in the wild}
\label{sec:usegen}
Passphrases can be generated by users or by computers (\emph{system-generated}). System-generated passphrases, %
uses an algorithm and generates pseudorandom passphrases based on a training corpus or wordlist. In this section, we will examine different in-use passphrase 
and analyze their properties.

\subsection{User passphrases}
\label{ssec:userpp}

To examine the class of user passphrases, we need to have a dataset compromising user passphrases. Unfortunately, unlike passwords, where data breaches are not uncommon, %
resulting in public dataets~\cite{fouriq}, there is no public dataset of passphrases available. So, we leverage a simple idea: password leak databases often contain long passwords that could potentially be passphrases without a proper delimiter. Thus we devise a segmentation algorithm to identify passphrases from password leak datasets and construct the first user-chosen, in-use passphrase dataset. %

\paragraph{Identifying user passphrase from leaked password data.} 
We leverage a compilation of prior password data breaches that surfaced in 2018 by 4iQ security firm~\cite{fouriq,pal2019beyond,li2019protocols}. \changed{The leaked dataset contains nearly 1.4~billion email-password pairs and has been used in prior works~\cite{pal2019beyond,sahin2021don,li2019protocols}}.

For our study, we only consider passwords that contain 20 or more ASCII characters (which is roughly 4 words given that the average length of an English word is 4.8~characters~\cite{Norvig}). There were 5.7~million such unique passwords.   Then, we filtered this list by removing passwords that are potentially hash values~\cite{pal2019beyond} --- containing only hexadecimal characters or follows popular hash formats~\cite{hashcat} --- or emails --- containing  `@' symbol in prefix and `.' in suffix and passwords less than nine English letters in them.   %
Then we segment passwords using segmentation algorithms. Segmentation algorithm tries to split the passwords into a sequence of meaningful words. We tried a hybrid segmention approach (combining two segmentation approaches) for segmenting passwords and effectively detecting user-generated passphrases (details in Appendix~\ref{sec:segmenting-pws}).

We found~$\nppws$ unique passphrases using our algorithm. %
The most frequent passphrase was used by 258 users, whereas 99.5\% of passphrases were used by only six or less users. %
We show the top three most used passphrases as well as three randomly sampled passphrases from the ones used by only one user in the top row of~\figref{fig:ex-passphrases}. We inserted `\textsf{-}' marks to show the segmentation. Finally, for checking the coverage of our method, we take 100 random passwords which are more than 20 characters but discarded by our algorithm.  we found 18 passwords that could be considered as passphrases but our algorithm failed to detect them. %
So our passphrase detection algorithm might miss out on passphrases that are heavily modified; but %
by construction, we have zero false-positives. We put more details on our method in Appendix~\ref{sec:inusepassphrase}. We will use these user generated passphrases (we will call it \userpp dataset) to empirically establish the guessability and memorability users achieve when they are choosing the passphrases themselves. \changed{We compared this passphrase dataset with leaked real-world password dataset to check for ecological validity. We found that linguistic properties of our dataset are inline with previous work on passphrases and frequency distribution of the passphrase dataset mirrors that of the original breached password dataset, hinting at ecological validity of our passphrase dataset. Detailed results are in Appendix~\ref{sec:ecolofical_passphrase}. }

\vspace{-1em}

\subsection{System-generated passphrases in use}
\label{ssec:systempp}

We focus on password managers to understand the in-use system-generated passphrase generation algorithms. We surveyed 13 popular password managers, such as LastPass, 1Password, KeePass~\cite{surveyList}. Among these, we found that three password managers provide a passphrases generation functionality: 1Password~\cite{1pass} and Enpass~\cite{enpass} use \dice, whereas Keepass~\cite{keepass} uses a template-based version of \dice (which we call \mmap) as their internal algorithm to generate passphrases for users. 
We describe \dice and \mmap below. %

\vspace{1mm}
\noindent \textbf{\dice .} The most commonly used system passphrase generation algorithm is \dice (called \dice from now on)~\cite{reinholddiceware, shay2012correct}, which relies on randomly choosing words from a dictionary and combining them to make a passphrase until the required length of passphrases is reached. Although \dice is highly secure, users find it very hard to remember the passphrases, i.e., the memorability of the passphrases is very low \cite{shay2012correct}. An in-use approach to improve upon the memorability of \dice passphrases is the \textit{template based \dice}~\cite{mmapCode} or \mmap.

\vspace{1mm}
\noindent \textbf{\mmap .}  This algorithm has a dictionary of English words segregated based on various parts of speech tags and has 27 syntactic templates for the English language, whose components are the tags, embedded within the algorithm \cite{mmapCode}. The idea of using syntactic templates has handled the issue of memorability very well. %
However, this approach has compromised on the security of the passphrases (\secref{ssec:guesseval}) and sacrificed the extensibility to generate arbitrary length strong passphrases as we will see next while evaluating the property of these passphrases. 

\subsection{Properties of in-use passphrases}
\label{ssec:langmodel}

In this section, we first demonstrate one defining characteristic of User passphrase is that it is closer to natural language than system-generated ones. \mmap is somewhat between \dice and \userpp in terms of being close to natural languages. %

\vspace{1mm}
\noindent \textbf{Linguistic properties.} We randomly sampled 3000 passphrases of similar length distribution from each system (dataset for \userpp) and compute their perplexity~\cite{brown-1992-perplexity} using GPT-2 model~\cite{radford2019language}. Lower perplexity indicates closer to natural language text. We noted that \userpp passphrases have a very low perplexity value similar to a natural language corpus, signifying a key reason for their memorability. On the other hand, the passphrases from \mmap is a close second in their resemblance to natural language, almost comparable to user passphrases, which indicates an improvement over the passphrases generated by \dice that performed poorly in this aspect. Details are in Appendix~\ref{sec:perplexity}.

\vspace{1mm}
\noindent \textbf{Issue with \mmap}: Although \mmap improved upon \dice, it comes with a compromise in security. On further investigation, we note two major shortcomings for \mmap. First, \mmap is not scalable as the information required by the system are all internally encoded within the algorithm \cite{mmapCode}. Any extension will potentially require a linguist to create new syntax rules and contextual wordlists (assuming it's possible for large passphrases). 
    Second, and more importantly, the security of the passphrases does not scale well with length. %
    We note that the guess ranks of these passphrases gets saturated around length 8---guess rank of 8-word passphrase is nearly the same as that as of 13-word. The potential reason is the constraint imposed by the underlying hardcoded and extremely limited syntax rule patterns of \mmap (Appendix~\ref{sec:mmap-shortcoming}). %

With these insights, we aim to improve the state of the art by balancing security and memorability of the passphrases. To that end, we design a novel constrained Markov Model-based optimization technique in~\secref{ssec:genmodel} for generating passphrases.

\newcommand{\tokenend}{END}
\section{\system: Optimizing guessability and memorability}
\label{sec:design}
To overcome the limitations of \userpp passphrases and system-generated passphrases, we design \system: an automated passphrases generation framework that can generate memorable as well as hard to guess passphrases. To do so, \system uses a constrained generative process by modifying a generative Markov model.  The constraints are based on approximate memorability and guessability metrics that we define.

\subsection{Memorability of passphrases}
\label{ssec:memorability} 

\noindent In order to improve memorability of system-generated passphrases, we first have to quantify memorability, and to that end, we use the notion of character error rate (CER), which is the rate of error per character while typing the text. Prior works~\cite{leiva2014representatively, mackenzie2002character, kristensson2012performance} noted that CER is widely accepted as a proxy for memorability.

Leiva et al.~\cite{leiva2014representatively}, however, tried to obtain memorable sentences which are representative of a corpus and are complete.  On the contrary, we want to ensure generating memorable passphrases which might not be complete and might not be representative of all passphrases users can memorize. Thus, the signals for memorability we want can be very different. %

In order to investigate, we used a dataset of 2,230 sentences, each of which has been annotated with the character error rate (CER) determined from a user survey~\cite{kristensson2012performance}. We calculated various parameters for each phrase in the dataset like frequency of occurrence, out of vocabulary words, the average frequency of occurrence, etc. With these data, we find the statistically significant correlation of each feature concerning CER (using pearson product moment correlation coefficient~\cite{lehmann2005testing}) and found three statically significantly correlated signals.

\paragraph{Unigram probability ($L_1$) of a phrase.} Calculated as the sum of the log of unigram probabilities of the individual words in the phrase. Phrases with a higher $L_1$ are likely to be more common, consisting of frequent words, and hence, easier to memorize \cite{freq}. In fact, $L_1$ has $p \approx 0$ (very high statistical significance) and $r=-0.83$ (very high negative correlation) with respect to CER.

\paragraph{Bigram probability ($L_2$) of consecutive words.} Similar to $L_1$, $L_2$ is calculated as the sum of the log of bigram probabilities of all the consecutive pairs of words in the phrase. Phrases with a higher $L_2$ are more likely to be closer to the natural language and thus it will be easier for a user to remember which is supported by its high statistical significance ($p \approx 0$) and high negative correlation ($r=-0.84$) with CER. 

\medskip
\noindent\textbf{Standard deviation ($\sdchr$) of the number of characters per word.}  Higher variability in the number of characters per word leads to higher processing effort and cognitive load. This is corroborated by the fact that $\sdchr$ has a very high statistical significance ($p<10^{-4}$) and positively correlated concerning CER ($r=0.25$).

Leiva et. al.~\cite{leiva2014representatively}, only considered unigrams to distinguish memorable sentences. %
However, in our work, we included bigram too as identified by our experiment. Higher bigram probability helps maintain syntactic structure, or the ``lexical distinctiveness'' to increase memorability~\cite{danescu2012you}.

Note that, computation of $L_1$ and $L_2$ requires a large corpus. In this work we use the \wiki dataset (\secref{sec:dataset}) for the purpose. The model was then fitted according to a generalized linear regression  with the above mentioned features. This yielded a good fit ($R^2 = 0.70$) for the CER estimate and we computed it for a passphrase $\pp$ as $\cer$($\pp$) = $\alpha_1\cdot\lprob$($\pp$) + $\alpha_2\cdot\lprobbi$($\pp$) + $\alpha_3\cdot\sdchr$($\pp$)
with $\alpha_{1} = -3.42\times 10^{-2}$, $\alpha_{2} = -6.46\times10^{-3}$ and $\alpha_{3} = 1.19\times10^{-4}$.

\subsection{Guessability of passphrases}
\label{ssec:guessability_old} 

\noindent As discussed in \secref{ssec:secandthreat}, the best way to measure the guessability (i.e., security) of a passphrase is to calculate its guess rank, which is, in fact, the estimated number of tries for an adversary to arrive at the correct passphrase. Recall that higher the guess rank, lower the guessability. To calculate the guess rank of each passphrase, we simulate any cracking algorithm with the support of suitable training data ~\cite{Kelley2012, Bonneau2012}. The cracking algorithm will then have a probability of generation associated with each model. We employ a Monte-Carlo simulation ~\cite{dell2015monte} that uses this probability to calculate the guess rank of the passphrase as discussed below.

In the pre-processing step, we generate $n$ passphrases $\{\beta_1, \ldots, \beta_n\}$ from the target model $\model$ along with their estimated probabilities of generation. We sort the probabilities in an array in  descending order as, $A=[\model(\beta_{1})....\model(\beta_{n})]$, and create the rank array $C$, where the $i$-th element is computed as:
\begin{equation}
    C[i]= 
    \begin{cases}
      \ceil{\frac{1}{n \cdot A[i]}}, & \text{if } i = 0 \\
      \ceil{\sum\limits_{j=1}^{i} \frac{1}{n\cdot A[j]}} = C[i-1] + \ceil{\frac{1}{n \cdot A[i]}},& \text{otherwise } \\
    \end{cases}
\end{equation}

A probability $A[i]$ corresponds to a rank $C[i]$, %
thus we estimate guess rank of a passphrase $\alpha$ with the largest $j$
such that $A[j] > \model(\alpha)$  through binary search. The  guess rank is estimated by taking an weighted average of the values of $C[j-1]$ and  $C[j]$. 

The above process allows us to calculate the guess rank 
from any automated cracking algorithm that can generate passphrases along with their probability of generation, which includes most of the current state of the art cracking algorithms ~\cite{dell2015monte}. But research has shown that a professional attacker using a semi-automated cracking process on a huge corpus of passwords can adapt to the dataset and thus is much more efficient than any known fully automated cracking algorithms \cite{realWorldAccuracies}. This idea can be extrapolated to the domain of passphrases and thus using any single model for estimating the guess rank will overestimate the number of guesses to arrive at the correct passphrases as compared to the number of guesses required by a professional adversary in practice.

To resolve this issue, we use the concept of \textit{min auto}, which has been briefly discussed in \secref{ssec:secandthreat}. Here, we take into consideration multiple models which can be used to estimate the guess rank of passphrases and have been trained in suitable training data. Ur et al.~\cite{realWorldAccuracies} have shown that for each password taking the minimum of all guess ranks estimated by the various models is a reasonable approximation to the actual guess rank needed in practice for passwords. Similarly, we extend the idea to passphrases by taking the minimum of $n-$gram word and character models ($2,3-$gram for words and $4,5,6-$gram for character) trained over a huge corpus of passphrases from the corresponding system to be evaluated. Some system-specific models have also been considered to obtain a more accurate approximation of the guess rank (\secref{evalsec}).
Leveraging this guessability and memorability framework we next examine the dataset \system uses for passphrase-generation.

\subsection{Curating a corpus for quantifying memorability and guessability}
\label{sec:dataset}

\noindent Our metric for measuring memorability in~\secref{ssec:memorability} requires a universal corpus and the \system needs a bigram Markov model for the generation of passphrases. We use Wikipedia data as a corpus of human-generated text data. %

\paragraph{Dataset.}   
We used a recent dump of Wikipedia articles\footnote{https://dumps.wikimedia.org/enwiki/latest/} and  used the Wikimedia Pageview API client to obtain the top 5\% articles based on the page view count, aggregated over the last five years. We then clean the data by removing all tags, URL links, and captions, and used case-folding~\cite{chatterjee2016password}. We also remove words with less than three characters, as well as numeric or alphanumeric words. Our final data contained 8,210 articles with over 29 million total words and more than 455 thousand unique words. We refer to this dataset as \wiki, and use it as a universal corpus for CER estimation throughout the paper and also use this corpus to generate
secure and memorable passphrases from \system.

\paragraph{Training bigram Markov model.} Next we trained a bigram Markov model on \wiki data.  We record all word-bigrams and  their frequency of appearance in the dataset. 
We will refer to this model as $\mkmodel$, %
and the log probabilities of unigram and bigrams in the dataset as $\lprob$ and $\lprobbi$. Thus $\lprob(\w) = \log({f_\w\over \sum_{\w}f_\w})$ and $\lprobbi(\w, w') = \log({f_{\w\w'}\over \sum_{\w''}f_{\w\w''}})$ and all logarithms are over base 10.  Note, we also assume $\mkmodel.\nextstr(\w)$ as a function that returns all the words that appear after $\w$ in the corpus. %

\subsection{Tuning guessability and memorability}
\label{ssec:tuning}

We estimate CER and guess rank using a bigram Markov model %
trained on the \wiki dataset in this work for estimating guess rank for the \system passphrases.

\paragraph{Optimizing memorability.}\label{memopt} Since we are trying to generate memorable passphrases, we focus on the syntactic structure, as well as the usage of simpler words, to help increase the memorability of the passphrase.  We thus introduce the generation probability of
a passphrase, based on the various parameters discussed in~\secref{ssec:memorability}. We express CER for a passphrase as $\cer(\pp) = \alpha_1 \cdot \lprob(\pp) + \alpha_2\cdot\lprobbi(\pp) + \alpha_3 \cdot \sdchr(\pp)$, and to generate memorable passphrases we aim to optimize $\cer(\pp)$ by controlling the parameters it depends on $L_1(\pp), L_2(\pp)$ and $\sdchr(\pp)$.
\par
\paragraph{Optimizing guessability.}\label{guessopt} We use the same \wiki corpus to train a bigram Markov model and also to generate the probability distribution of every unigram and bigram the \system uses for generation of passphrases. We then use the Markov model as one of the algorithms used in the estimation of guess rank of the passphrases generated by \system as shown in Section \ref{ssec:guessability_old}. We use the unigram and bigram probabilities for calculating the $\lprob$ and $\lprobbi$ of passphrases for the estimation of CER.

\subsection{Generative model}
\label{ssec:genmodel}

Once we have curated our corpus, and the metrics have been suitably defined, we start generating passphrases. %
Recall that in our case CER is heavily dependent (and linear combination) of its factors. The equation obtained previously for the CER of a passphrase (using \wiki dataset), helps us have a better estimate to optimize our generated sentences. For ensuring high guess rank too, we try to maintain a trade-off between these two metrics to generate syntactic and secure passphrases using a simple idea: high unigram/bigram probabilities ensure high memorability and low guess rank, so choosing the right words with optimum probability might ensure both memorability and security.

\paragraph{Generation.} We start generating passphrases based on the current word. In subsequent words, we evaluate the whole support based on thresholds. For every state change, we recheck our CER and guess rank estimates to obtain a reasonable choice of word.  For the successful generation of passphrases, we pass the trained Markov model to the \system along with the desired length of a generation, $L$, to generate passphrases adhering to our constraints as discussed below.

We begin with the start token \start, and the first word appended to the string is from the list of words a sentence begins within the corpus provided that is not a stop word. \textit{Stop words} are a set of commonly used words in any language. Some examples in English - the, are, and, over, etc. We do this to ensure less predictability in our passphrases as there is a high probability of the generated passphrase starting with a stop word otherwise.\par

We use a score function modeled on the observation that an approximate CER can be computed incrementally. That is, given a partially generated passphrase $\pp_i = \w_1\ldots\w_i$, one can compute the intermediate CER value using the following equation.  
\begin{equation*}
  \score(w_1\ldots w_{i}) = \alpha_{1}\lprob(w_{i})+\alpha_{2}\lprobbi(w_{i-1},w_{i})
  +\alpha_{3}\sdchr(w_1\ldots w_{i})
\end{equation*}

Similarly, we also know that the guess rank of the generated passphrases is dependent on the bigram probabilities $\lprobbi$ when the Markov model trained on the \wiki corpus is used as the cracking algorithm. Thus we use the following constraint while generating passphrases: $\score(\w_1\ldots \w_i) \le \thetaone$ 
and $\lprobbi(\w_{i-1}, \w_{i}) \le \thetatwo$, where $\thetaone$ and $\thetatwo$ are two system parameters.\vspace{6pt}

Relaxing thresholds on $thetaone$ and $thetatwo$ a lot will make the quality of generation similar to that of a normal Markov model and tightening them might result in a reduced sample space. Keeping this in mind, the fractions can be tinkered around as per need. Even though they have a common factor, $\lprobbi$, between each other, the optimization can be considered independent. %
Since the flow of a sentence is based on qualitative inspection, the thresholds depend on the corpus itself, and thus, can vary depending on the user's need. For example, a corpus with a higher percentage of rare bigrams will automatically generate less memorable sentences, thus requiring us to relax the upper bound for CER and vice versa. We provide a detailed discussion on trade-off and bound of thresholds on $\thetaone$ and $\thetatwo$ in Appendix~\ref{sec:thresholdtradeoff}.

\newcommand{\getfirstword}{\mathsf{GetFirstWord}}

\begin{figure}[t]
  \centering
  \fpage{0.45}{
  \fontfamily{cmss}
    \underline{$\getfirstword(\mkmodel):$}\\[2pt]
    $\PW \gets \mkmodel.\nextstr(\start)\setminus \sw$\\
    $\w\genfrom{\lprob}\PW$\\
    return $\w$\\[5pt]
    \underline{$\sysgen(\pwlen, \mkmodel)$:}\\[2pt]
    $\w_1\gets \getfirstword(\mkmodel)$\\
    $i \gets 2$\\
    while $i\le \pwlen$ do\\
    \mytab $\PW' \gets\mkmodel.\nextstr(\w_{i-1})$\\
    \mytab if $i = \pwlen$ then\\
    \mytab \mytab $\PW' \gets \PW' \setminus \sw$ \commentNew{Remove stopwords}\\
    \mytab \commentNew{CER and guess rank constraint}\\ 
    \mytab $\PW' \gets \{\w\in\PW' \given \score\left(\w_1\ldots\w\right) \le \theta_1$ and $\lprobbi(\w_{i-1},\w) \le \thetatwo\}$\\
    \mytab if $i=\pwlen$ then \commentNew{Ends in a end symbol}\\
    \mytab \mytab $\PW' \gets \{\w\in\PW' \given \lastword \in \mkmodel.\nextstr(\w) \}$\\
    \mytab $\w_{i} \getsr \PW'$\\
    \mytab if $\w_{i} = \bot$ then \\
    \mytab \mytab$\w_1 \gets  \getfirstword(\mkmodel)$ \commentNew{No passphrase found; restart} \\
    \mytab \mytab $i \gets 1$\\
    \mytab\ $i\gets i$+$1$\\
  return $\w_1\ldots\w_\pwlen$
}
\caption{The \system algorithm. The algorithm generates a passphrase of length $\pwlen$ given a bigram Markov model $\mkmodel$. Here $\sw$ is a set of stop words, and $\start$ and $\lastword$ are the start and end symbols used in the Markov model. Here $\genfrom{\lprob}\PW$ denotes sampling from the support $\PW$ but according to the probability distribution assigned by unigram probabilities (without log); similarly
$\getsr \PW$ denote sampling uniformly randomly from the elements in $\PW$.
}
\label{fig:algorithm}
\vspace{6pt}
\end{figure}

\subsection{The final algorithm, \system{}}
\label{sec:finalalgo}

We introduce \system{}, a step-by-step approach to generate passphrases, under constraints of memorability and guessability, while preserving its syntax and meaning in ~\figref{fig:algorithm}. The actual implementation is an optimized version of the one shown, where we use $O(\log n)$ time for the sampling of the next word, where $n$ is the number of choices, as opposed to the $O(n)$ shown in ~\figref{fig:algorithm}, with the help of some pre-processing.

The algorithm is greedy, and not necessarily optimal. But, replacing the corpus or changing any of the variables can essentially just be a straight swap with the existing one, based on user preference or need, making the algorithmic approach a generalized version of generating optimized passphrases.

\paragraph{Selecting $\thetaone$ and $\thetatwo$.} We try to ensure that \system{} favors rarer bigrams (low $\thetatwo$) while maintaining a low intermediate CER score (low $\thetaone$). We also note that $\score(\cdot)$ function has $\lprobbi$ in it, so we can bound the value of $\score$ given $\thetatwo$. That is to say, if $\lprobbi(\w_{i-1}, \w_i) \le \thetatwo$, then $\thetaone \ge \score(w_1\ldots w_i) \ge \alpha_1\lprob(\w_i) + \alpha_2\thetatwo + \alpha_3\sdchr(\w_1\ldots\w_i)$, because $\alpha_2$ is negative. Thus, for a given $\thetaone$, the value of $\thetatwo$ can be
bounded. $\sdchr$ and $\lprob$ is at least 0, then $\thetaone \ge \alpha_2\thetatwo$, or $0 \ge \thetatwo \ge {\thetaone\over\alpha_2}$.

We chose $\thetatwo$ to be at 80\% of the minimum possible value of $\lprobbi$ (giving the system a leeway of within 20\% of the minimum value), which is -17.4. This will ensure that the generated passphrases contain rare bigrams and thereby high guess rank.  Setting $\thetatwo= 0.8 \times -17.4$, gives us $\thetaone \le 0.5$, as $\alpha_2=-0.00646$. We use these values for generation in the design of~\system.

Our generation is incremental ensuring the invariant of CER (memorability) and guess rank (guessability). An alternative approach would have been generating the whole passphrase of length $\pwlen$, and then checking if the CER and guess rank constraints are met. Intuitively, we can see that the current approach is much more efficient than the alternative and can generate usable passphrases a lot quicker. %
This is further demonstrated below.

\paragraph{Execution time. }\label{sssec:exectime} To evaluate the performance of the final \system algorithm, we compare it with multiple variations and baselines. The baselines we take into consideration are \dice, \mmap, as well as a basic Markov model trained on the \wiki dataset. We also examine the variation of \system where all rejections (according to constraints of $\thetaone$ and $\thetatwo$) take place after the entire passphrase is generated by a Markov model (\system$_{\text{end}}$). This variation can be understood as a system that can create optimal passphrases for the constraints imposed. For reasonable comparisons, we keep the rest of the system parameters the same for both these versions.

After generating 1000 passphrases of equal length distribution across all the systems taken into consideration, we checked their execution time, which includes the pre-processing time as well. \dice, \mmap, and Markov run in 7.85 seconds, 0.33 seconds and 0.05 seconds, respectively. In comparison, \system takes only 0.08 seconds, which is even comparable to a generic Markov model, and much better due to the guarantees offered by the generated passphrases on their memorability and security. Looking further, the execution of the model that can generate the most optimal passphrases for the set constraints, \system$_{\text{end}}$, takes a very significantly larger 810 seconds to complete.%

\section{Evaluating passphrases}
\label{evalsec}

\noindent Our goal is to improve upon the passphrases generated by template-based diceware by resolving its shortcomings while still not losing out on the advantages it offers. In other words, we would like to generate passphrases that are easier to remember while still being hard to guess. In this section, we will evaluate the quality of passphrases generated by \system and compare it with passphrases used by users or generated using other methods---we will compare the following five sets of passphrases:

\textbf{\dice}. \dice was proposed earlier for generating passphrases by random selecting a sequence of words from a vocabulary~\cite{reinholddiceware}. We use a wordlist commonly used as vocabulary \dice~\cite{bonneau_2018}. These passphrases are in general much harder to guess (e.g., ``clay reactive smasher authentic chrome hamster'').

\textbf{\mmap}. An improved version of \dice, where passphrases are generated based on predefined syntactic templates for the English language. The templates are composed of various parts of speech like nouns, verbs, adjectives, etc., which will be replaced with suitable words from a vocabulary segregated in a similar way \cite{mmapCode}. The passphrases generated in such a way are relatively easier to remember (e.g., ``when does a bellboy spike an elect but not a sidebar'').  

\textbf{\markov}. We also use a bigram Markov model trained on the \wiki dataset as a baseline for comparison considering that \system is an enhanced version of the former. The process of passphrase generation is similar to \system. However, we don't impose any constraints on the intermediate steps and sample words weighted on their conditional bigram probability (e.g., ``leopold arranged for some users include the war'').

\textbf{\userpp}. We identified several user-created passphrases from prior password leaks ( in~\secref{sec:usegen}). These passphrases are user-created, close to natural text and therefore, should be very easy to remember (e.g., ``just another happy ending''). 

\textbf{\ours}. Finally, we consider the model we propose --- \system~ --- which also internally uses a bigram Markov model trained on the \wiki dataset, with several control parameters to ensure the generated passphrase has higher memorability while maintaining a high guessrank (e.g., ``edge bands influenced how far north south''). Later we detail training and passphrase-generation of \system (\secref{ssec:genmodel}~and~\ref{ssec:constraints}).

We add a few random samples of passphrases from each set in Appendix~\ref{sec:passsample}. We compare the memorability of these five sets of passphrases using CER and strength using guessrank.

\vspace{2mm}
\noindent \textbf{Test sample.} We generated one million passphrases from template-based diceware, and following the same distribution of lengths, we generated same number of passphrases from each of \dice, \markov and \system. We use these as test sample in our evaluations. For \userpp, owing to the limited size of the dataset, we used only 6,500 user passphrases as our test sample. %
Note that the length distribution of \userpp passphrases is different from others, as users often tend to utilize passphrases of smaller lengths. We did not use the length distribution of \userpp for the system generated samples for this reason---as that would bias the test samples from \dice, \markov and \system towards smaller length passphrases.

\subsection{Strength of the passphrases}
\label{ssec:guesseval}

We measure the strength of a passphrase based on their guessrank %
using the \textit{min auto} approach discussed in~\secref{ssec:guessability_old}. \changed{Note that we considered an offline generalized untargeted adversary as mentioned in~\secref{ssec:secandthreat}}. To compute a min-auto rank of password we take the minimum guessrank according to a number of guessrank estimations. Prior work has shown such approach provides close approximation of
real-worlds. 

We considered seven guessing algorithms for the min-auto approach: 2-gram and 3-gram word-based Markov models and  4-gram, 5-gram, 6-gram character-based  Markov model~\cite{markovProb}, Wiki-5 bigram model, and template-based guessing algorithm. We explain  last two guessing models below. 

\textbf{Probabilistic \wiki bigram Markov.} For \markov and \ours, other than training on a huge corpus of generated passphrases, an attacker can also train it on the dataset using which the passphrases are generated, namely \wiki. As a bigram Markov model is used for the generation of the passphrases in these two systems, we also use a probabilistic bigram Markov guessing model trained on the \wiki dataset. The rest of the guessrank estimation is similar to the process described above.

\textbf{Template based estimation.} An attacker can use the fact that the passphrases generated by \mmap are finitely bounded by the templates that are being used. Thus, a guessrank for a passphrase generated by the algorithm can be guessed by trying out all the possible passphrases across all templates. To simulate this process, we first find the number of passphrases each template can produce. We then randomly choose templates one by one until the source template for that passphrase is chosen (we remember the source template so that we can estimate the guessrank, it is not available to the attacker) and count all the passphrases that the attacker would have enumerated by then, which will give us the guessrank for that passphrase. This guessing strategy is particularly effective against \mmap, and limits the largest guessrank a \mmap generated passphrase can achieve.

Recall that according to the threat model described in \secref{ssec:secandthreat}, the attacker has access to a huge corpus of passphrases generated from the system whose passphrases are being cracked. Therefore, the attacker have $10^7$ system-generated passphrases from \{\dice, \markov, \ours, \mmap\}, on which the attacker can train his cracking algorithm. For \userpp passphrases, the attacker trains on a smaller set of 70 thousand passphrases. 

This trained model is then used to guess the passphrases of the corresponding test samples. Each passphrase has a probability of generation according to a model and using the method in \secref{ssec:guessability_old}, we can estimate the guessrank that is, the number of guesses the attacker will need to guess the passphrase correctly using that particular guessing model. A smoothing factor is used for any out of vocabulary (OOV) n-grams encountered. We took the minimum guessrank across all the models as the final output.

\begin{figure}[t]
  \includegraphics[width=6.5cm]{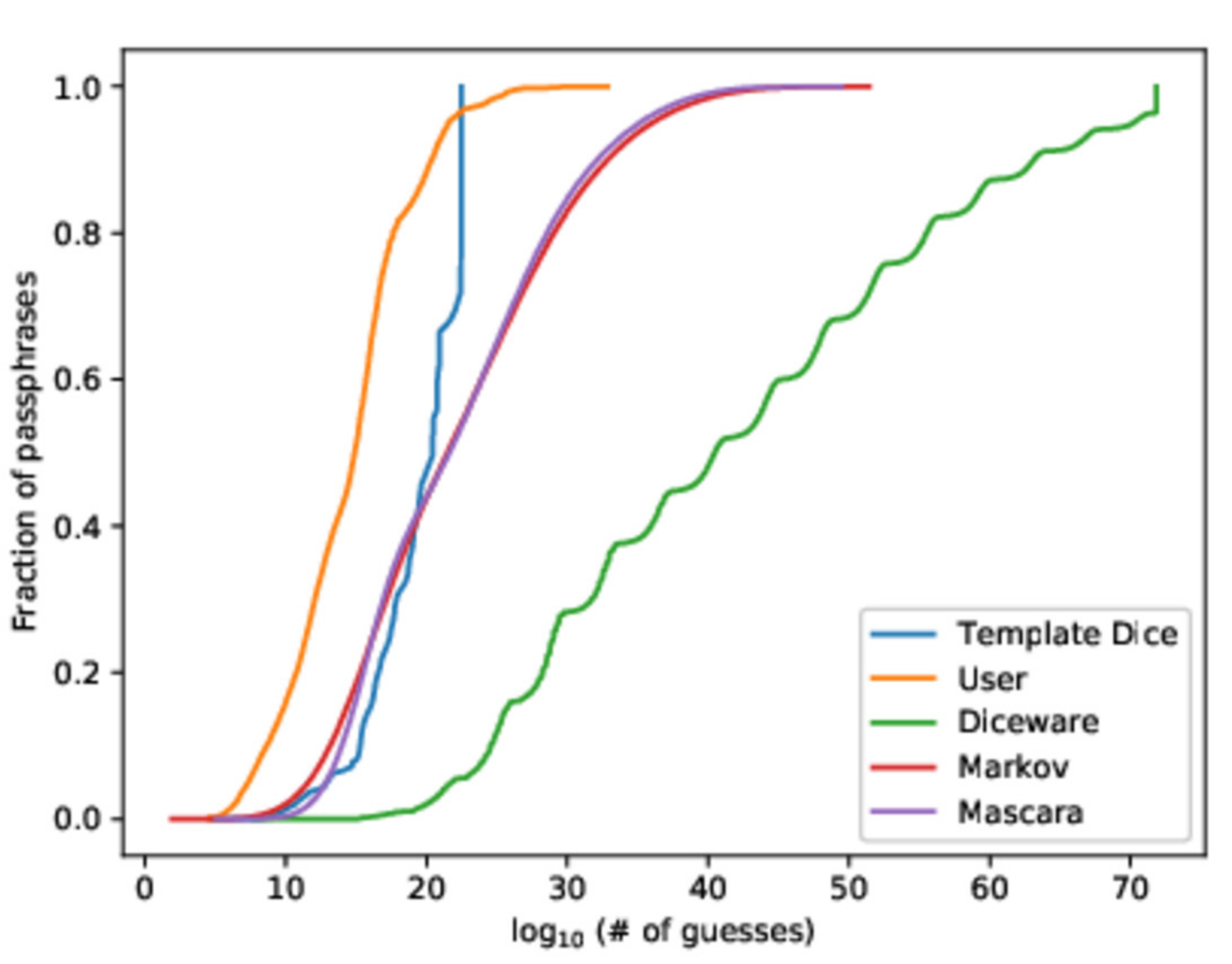}

  \caption{Comparing the strength of different sets of passphrases by evaluating their guessrank. Here, a cumulative distribution frequency of log guessrank (base 10) is shown. We can see that diceware passphrases are the most secure, while the user ones being the most predictable.
  }
  \label{fig:gr-all}
\end{figure}

\textbf{Results.} We show the (estimated) guessranks of the passphrases in the test samples in ~\figref{fig:gr-all}. \dice passphrases are the most secure with 50\% of passphrases requiring at least $10^{40}$ guesses. On the other end of the spectrum, we have \userpp passphrases, with 50\% of passphrases guessed within $10^{14}$ guesses, which is not even as secure as some of the most secure passwords \cite{realWorldAccuracies}. The predictability of \userpp passphrases can be somewhat attributed to their smaller length, but since that is the inherent nature of these passphrases, we did not see fit to change it. In between \userpp and \dice, we have \ours, \markov and \mmap. The security of \markov and \ours are similar, with \ours having a slight advantage over \markov in the 20\% most predictable passphrases of each set.

The main aim of \ours is to resolve the shortcoming of \mmap. Comparing these two, we see that the only advantage the latter has over the former is that the guessrank of \mmap is slightly higher than \ours for passphrases below the 40$^{\text{th}}$ percentile. These are the passphrases of a length less than 8. As we move along the curve, we observe a huge difference in the number of guesses needed by \ours and \mmap for their most secure passphrases. Template-based diceware needs $10^{22}$ guesses for at least 20\% of the passphrases, whereas \ours significantly improves upon it and requires over $10^{30}$ guesses. %

However, we argue that memorability of passphrases is another important criterion that should be considered while picking a passphrase. We discuss the memorability of passphrases next.

\subsection{Memorability of passphrases}
\label{ssec:memeval}

\noindent Passphrases must be memorable while being difficult to guess to be usable in practice. In this section, we measure the memorability of the passphrases in the test samples based on the character error rate (CER) estimate we devised in ~\secref{ssec:memorability}. CER estimates the probability of making an error while typing from memory (and not the actual \#characters that one might get wrong).

\begin{figure}[t]
  \includegraphics[width=6.5cm]{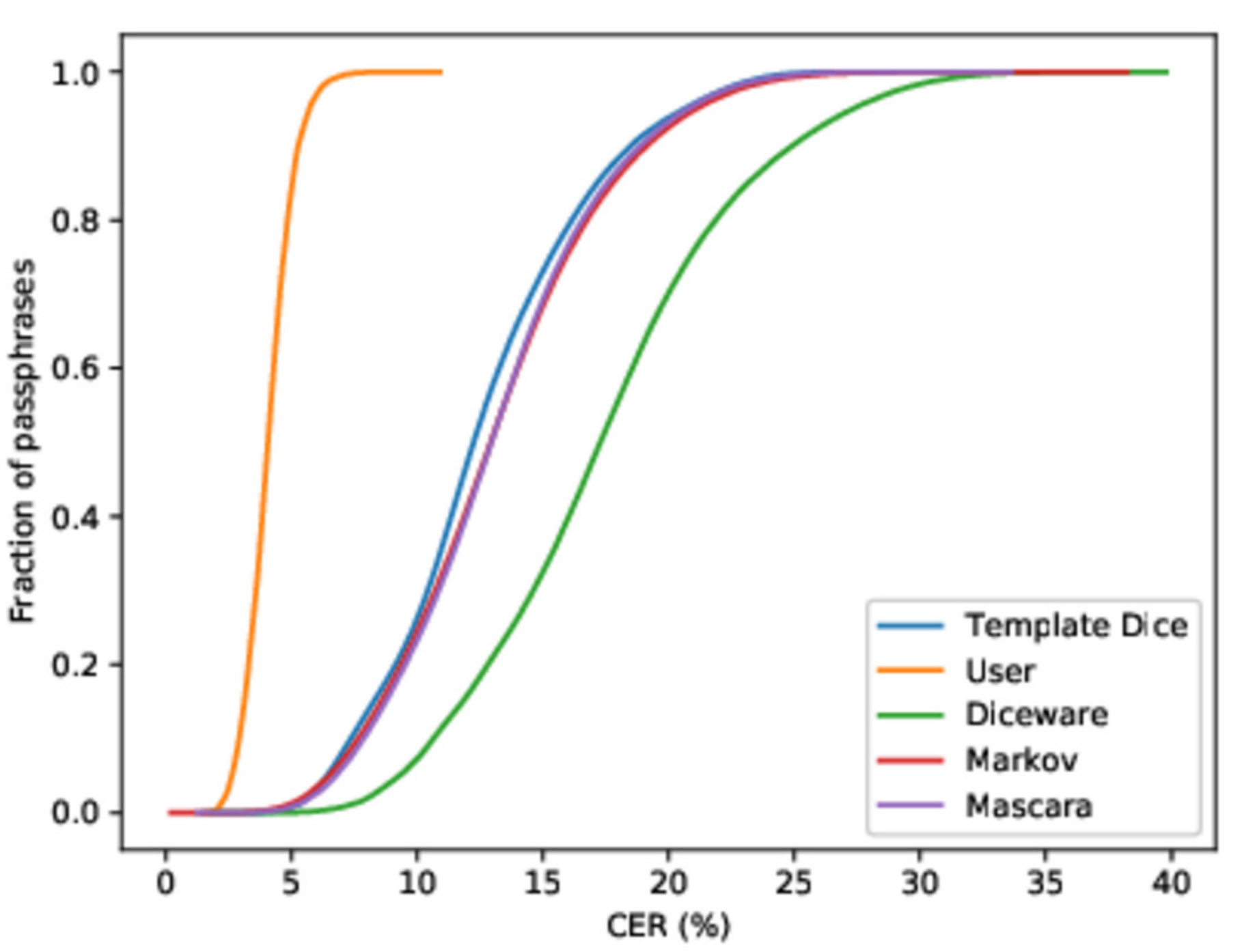}

  \caption{Distribution (CDF) of CER (Character Error Rate) for passphrases among the various samples. \dice passphrases have the highest CER, making it the least memorable and \userpp passphrases have the lowest CER which in turn means that they are easiest to remember.}
  \label{fig:cer-new}
\end{figure}

\paragraph{Results.}  The distribution of CER of the passphrases are shown in ~\figref{fig:cer-new}. As expected, \dice passphrases have a very high CER --- 50\% passphrases have CER of more than 17\% --- meaning that users are likely to make a mistake every 6 characters they type, which indicates a very low memorability. We hypothesize the lack of any syntactic structure is responsible for such high CER.  \userpp passphrases seem to perform the best, with 80\% of the passphrases with less than 5\% CER (a mistake every 20 characters). This is within expectations, given that users, in general, choose highly common phrases, quotes, and song or movie titles.

CER values of passphrases from \ours, \markov and \mmap are between \userpp and \dice. All the three CERs are almost similar (with \mmap having a slight advantage), with each of them having a 12.5\% CER for at least 50\% of the passphrases in their corresponding samples--- significantly better than \dice, although much worse than \userpp passphrases.

\paragraph{Takeaway.} We measure the guessrank of the passphrases using the \textit{min auto} approach employing multiple models to estimate the guessrank close to what a practical adversary might achieve. The memorability of the passphrases is determined by a commonly accepted proxy, CER.

Although we would like to increase the guessrank (decrease the guessability) of passphrases while reducing the CER, they are inversely correlated (intuitively, more structure to passphrases leads to high predictability) and thus one cannot be lowered without increasing the other. We tried to find a sweet spot that allows us to increase as much guessing resistance as possible while keeping the CER values close to what user-chosen passphrases enjoy. 

The results show that \system mitigate  the limitations of \mmap discussed in \secref{ssec:langmodel} and is potentially much more effective to use in practice. Although \dice offers significantly high security, users will also find it hard to remember, owing to its high CER. On the other end of the spectrum, while \userpp passphrases are very easy to remember, they offer little to no security. Furthermore, given the parameterized generation process of \system, one can
configure it to a setting that meets their needs, giving room to significant personalization.

\section{User Study}
\label{sec:workerstudy}

\noindent We used CER to estimate the difficulty of memorizing various passphrases (in~\secref{ssec:memeval}) analytically. However, to evaluate if indeed,  final \system{} generated passphrases are memorable to the users we ran a user study. %

\paragraph{Ethical considerations.} As the primary authors' institution did not have an official ethics review board, we did not obtain any official ethics review of the study. However, we discussed extensively with our peers and followed best practices of ethical research (e.g., principles set by Belmont Report ~\cite{beauchamp2008belmont}). The study was performed with informed consent. We asked for no sensitive (e.g., actual in-use passphrases) or personally identifiable information (such as email id).  We also removed the worker ids from survey results during analysis, reducing  privacy risk to the participants.

\begin{figure}[t]
  \centering
  \small
  \begin{tabular}[t]{lccc}
    \toprule
    Model & Part1 count & Part2 count & Return rate\\
    \midrule
    \ours & 72 & 61 & 84.72\% \\
    \mmap & 78 & 63 & 80.76\% \\
    \markov & 69 & 41 & 59.42\% \\
    \dice & 82 & 50 & 60.97\% \\
    \bottomrule
  \end{tabular}
  \caption{\#participants who completed Part 1 and returned for Part 2 with return rate (across different algorithms).}
  \label{fig:parti_count}
\end{figure}

\subsection{Design and setup}
\label{ssec:design}

Our two-part survey-based user study was deployed on crowdsourcing platform Prolific, where we assigned Prolific participants at random to one of passphrase generation algorithms out of \mmap, \ours, \markov, \dice and show them a passphrase to remember. For each algorithm, we randomly generated passphrases while uniformly choosing passphrase length from one of three length ranges ($\le$ 7, 8-12, >12). We used the range of $\le 7$ as most of user-generated passphrase (in our previous dataset from \secref{ssec:langmodel}) lies in the range. We recruited participants on Prolific with a greater than 99\% approval rate for our two-part survey. We selected US Citizens who are fluent in English and are above the age of 18. All the approved participants from part 1 of the survey were invited to carry out part 2. Our study was designed based on prior work on measuring memorability of login credentials~\cite{woo2020we}.

\textbf{Part 1.} In the first part, we asked each participant to choose from the three passphrases shown to them, all of which were generated by the same algorithm (randomly chosen for the participant) and were of the same length. %
Participants were asked not to write down or copy their chosen passphrase, and to only rely on memory. They were made to practice their chosen passphrase five times after finalizing their choice. Then, each participant was asked to authenticate twice---at the end of part 1 of the survey, and the beginning of part 2. We allowed at most five tries to authenticate in both parts of the survey. The users were asked not to paste their answers and were also assured that they would receive payment regardless of their authentication success. To distract participants before asking for authentication, the participants answered demographics, some generic questions and attention check questions. %

\textbf{Part 2.}  \changed{We invited participants who successfully finished part 1 to return and authenticate again after 48 hours from their completion time.} %
A two-day recall time interval is used and justified for most prior password and passphrase recall research~\cite{woo2020we,Kelley2012,joudaki2018reinforcing,huh2015surpass}. In fact Huh et al. argued that a good recall rate after two days potentially (empirically) signifies the practically required memorability of passphrase---thus we used the same interval~\cite{huh2015surpass}. 
\changed{However, 24\% of the users responded to the second part of the survey post 96 hours (4 days or more).}
At the end of the authentication, we provided a short survey to assess participants' perception of the chosen passphrase and how they may want to modify the passphrase. 

\changed{We paid \$0.75 to the participants who completed part 1 and \$1.00 for the participants who returned and finished their authentication in part 2. The participants took a total of 10 minutes on average to complete both parts.} The survey instrument is in Appendix~\ref{sec:userstudy}. For analysis, we used Kruskal-Wallis (KW) test and pairwise Mann-Whitney U tests to find statistically significant differences at $\alpha=0.05$~\cite{lehmann2005testing}. 
\changed{Our registration and login prompts are designed to reflect a real login system in line with a plethora of previous work~\cite{egelman2013ecologicalpass, woo2020we, ur-measure}. Like earlier work on conducting ecologically valid password study~\cite{egelman2013ecologicalpass}, we did not nudge the user to choose/remember passphrases for any specific system (e.g., \dice or \ours). Thus, we strongly believe our data (e.g., recall) also captures users' ecologically valid preferences.}

\subsection{Demographics and participants info}
\label{ssec:demograph}

A total of 310 Prolific users participated in our user study (we also ran 5 pilots in the survey development phase and updated questions to remove ambiguity according to pilot feedback). Among these participants, we detected invalid responses from 9 participants (2.9\%) who failed the attention check. After excluding those, 301 participants successfully chose the passphrases, practiced them, and answered all the survey questions properly. 

Forty-eight hours after Part 1, we emailed participants to return for the authentication in the second phase. Out of 301 participants, 217 participants returned within 24 hours of sending email, yielding a return rate of 72.1\%. Out of these participants, 202 (93.1\%) self-reported that their desired passphrase length (in number of words) is 7 or less for being able to remember in daily use. 

Of the 301 users, 70\% and 24.2\% of them reported their gender as female and male, respectively, while the rest either reported it as non-binary or they did not prefer to reveal it. Among the users who participated, the two highest age groups reported were 18-30 (53.16\%) and 31-45 (30.56\%). Also, 36\%, 29\% and 17\% of the users reported their highest qualifications as Bachelor's Degree, Some college, Master's Degree, respectively. We found no statistically significant differences across the models in their gender, age group, or highest qualification. Only 13\% participants reported working in or having education in IT or related fields.

\paragraph{Return rate is high for \ours{} due to better recall:} The distribution of our  participants across the different models and their return rate is also shown in \figref{fig:parti_count} (demographics are similar). The return rate is very high in \ours, almost 25\% higher than \dice. To further investigate, we checked the Part 1 recall rates (posted right after practicing 5 times) between the users who returned in part 2 and those who did not. We omit the detailed results for brevity, but we found that across all algorithms, users who did not return in part 2 has a significantly low part 1 recall rate. This difference is more prominent for passphrases with length 7 or less. E.g., for that passphrase length \ours have a recall rate of 92.31\% in ones who returned for part 2 and 60\% for those who did not. These results hint that the low return rate for \dice and \markov is indicative of the underlying fact that a significant fraction cannot recall the passphrases. In fact one participant mentioned for a Dice passphrase that \textit{``It was the longest and used random words''} and another mentioned for a Markov passphrase that \textit{``I'm already pretty confident I'm not gonna remember this one guys :(''}. These results indicate that some \dice and \markov passphrases are hard for users to remember.

\subsection{Passphrase statistics}
\label{ssec:pass_stats}

\noindent \textbf{Word length.} To make sure the statistical analysis yields a proper comparison, we would like the distribution of the word lengths of the passphrases across the different models to be similar. The average word length across the passphrases chosen in part 1 among \ours, \mmap, \markov, and \dice are 9.19, 8.74, 8.62, and 9.06, respectively. We perform a KW test to see how different the underlying distributions are. The KW test fails to reject the null hypothesis ($H = 3.59, p = 0.31$), confirming that the underlying distribution is not significantly different. The distribution of the word lengths of the passphrases across returning participants for part 2 among \ours, \mmap, \markov, and \dice are 9.42, 8.81, 8.85, and 9.16, respectively. Similar to part 1, the KW test gives $p = 0.39$ ($H = 2.97$), identifying that the underlying distribution is not significantly different among part 2 participants.

{\bf Passphrase strength.} We estimate the strength of the passphrases by their guess rank, which was calculated apriori using the process mentioned in \secref{ssec:guesseval}. %
The mean log guess rank (base 10) across \ours, \mmap, \markov, and \dice are 14.80, 11.70, 14.46, and 36.49, respectively. Similarly, the median log guess rank (base 10) across \ours, \mmap, \markov, and \dice are 14.20, 12.51, 12.86, and 36.05, respectively. %
Thus, for non statistically different length distributions, the \dice and \mmap are most and least secure, respectively.

To compare the underlying distribution of the guess rank among the four algorithms, we again perform the KW test. This KW test rejects the null hypothesis ($H = 157.7, p \approx 0)$, confirming that the underlying distributions are statistically significantly different. We then perform the Mann-Whitney U test on all possible pairs, and further find that guess ranks of \textit{each pair} to be statistically significantly different. Next, we check the memorability of these passphrases using survey responses. 

\subsection{Evaluating passphrase memorability}
\label{ssec:recall_cer}

\begin{figure}[t]
  \centering
  \small
  \begin{tabular}[t]{lccc}
    \toprule
    Model & Recall & Mean CER & Median CER \\
    \midrule
    \ours & 26.23\% & 34.78\% & 35.85\% \\
    \mmap & 17.46\% & 35.44\% & 36.58\% \\
    \markov & 21.95\% & 37.84\% & 41.27\% \\
    \dice & 24.00\% & 38.49\% & 42.57\% \\
    \bottomrule
  \end{tabular}
  \caption{\% participants with successful recall, mean and median CER (as \%) while authenticating after 2 days.\ours perform best. Surprisingly \dice is close second.}
  \label{fig:recall_cer}
\end{figure}

\begin{figure}[t]
  \centering
  \small
  \begin{tabular}[t]{lccc}
    \toprule
    Model & Recall & Mean CER & Median CER \\
    \midrule
    \ours & 46.15\% & 19.10\% & 8.82\% \\
    \mmap & 28.57\% & 30.03\% & 34.48\% \\
    \markov & 14.29\% & 39.74\% & 43.59\% \\
    \dice & 23.08\% & 45.48\% & 56.41\% \\
    \bottomrule
  \end{tabular}
  \caption{\% participants who were shown passphrase length $\le$ 7, with successful recall, mean and median CER (as \%) while authenticating after 2 days. Recall of \ours is 2x of that of \dice and 1.6x of \mmap.%
  }
  \label{fig:recall_cer_less7}
\end{figure}

\paragraph{Recall and CER.} In our case, \textit{successful recall} and \textit{low CER} signify high memorability. Recall is successful if in part 2 users correctly input every character (including spaces) of the passphrase within five attempts. Also, the character error rate (CER) for a particular attempt is the edit distance between the attempt and the original passphrase divided by the number of characters in the original passphrase. We calculate CER for a user as the minimum CER across all attempts.
In part 1 (right after seeing the passphrase) \ours, \mmap, \markov, and \dice have  similar recall rates (between 50\% to 60\%) overall. However, for passphrases of length 7 or less (self-reported as preferred by more than 90\% participants), \ours has a Part 1 recall rate of 83.3\%, significantly outperforming other algorithms (60\%---65\% for this length range).

\textbf{Recall after two days.} \figref{fig:recall_cer} shows the percentage of users who were able to successfully recall passphrases chosen in part 1 (recall rate), as well as the mean and median CER for all users after 2 days. \ours{} has the highest recall rate of 26.23\% among all algorithms (\mmap recall only 17.46\%). Surprisingly, the recall rate of \dice is 24\%, not too far from \ours. %

However, when we investigated further, we made two key observations. First, the return rate of \ours was 84.72\% and that of \dice was only 60.97\%. As we discussed before, the recall rate in part 1 for users who did not return in part 2 was significantly lower than the ones who returned. So, we already have close to 40\% users from part 1 who did not remember \dice passphrase (as opposed to less than 16\% for \ours). 

Second, for participants who returned for part 2, \figref{fig:recall_cer_less7} shows the recall rate, mean and median CER for passphrases of length 7 or less, which the majority of participants wanted to use. We hypothesize that for remembering higher-length passphrases, there are other confounding factors including strong user bias against using very long phrases for day to day use. For passphrases of length 7 of less, \ours have a recall rate of 46.15\% after two days of no forced practice and less than 10\% median CER. This recall rate is 2x higher than \dice and 1.6x higher than \mmap, which are used in real-world systems. Furthermore, the median character error rate (CER) for \ours passphrases (8.82\%) is 6-times lower than \dice (56.41\%) when the passphrases of length seven or less are considered. \changed{We further checked if the edit distances between typed passphrases and actual passphrases for \ours is statistically significantly lesser than other algorithms (lesser edit distance impliers lesser error and more memorability). We ran pairwise Mann-whitney U tests (with Bonferroni correction for multiple tests) over the edit distances for these passphrases in part 2. The edit distances for \ours passphrases is indeed statistically significantly lower than other algorithms (p < 0.005). In fact the average edit distance for \ours is 6.27 at the end of part 2 which is approximately 2.9 times less than \dice (18.61)}. So, for the desired length of passphrases (as self-reported by the users) \ours significantly improves the memorability of state of the art passphrase generation algorithms.

\changed{\system{} is not a human-in-the-loop (HITL) approach. However, some prior work took a HITL approach of passphrase generation too. Just to test the utility of \system{} (even though it's not a fair comparison) we compare \system{} with a guided word choice (GWC) method by Blanchard et al.~\cite{blanchard-2018-gwc} using a separate user study (Appendix \ref{sec:gwc}). Although GWC provides users more control for choosing their random words, the user study reveals that Mascara-passphrases are still at least comparable or a little more memorable (and less error prone) than GWC-generated passphrases.}

\section{Conclusion}
\label{sec:conclu}
\noindent In this work, we take the first step towards the systematic generation of memorable yet secure passphrases. We presented a novel in-use passphrase data-set and leveraged the linguistic properties to propose~\system{}, a passphrase generation method. In the process, we also created a framework for measuring memorability and guessability of passphrases. Our exploration, aside from creation of \system{}, provides multiple important insights. First, our ecologically valid in use  \userpp passphrases show that users, left to their device choose weak passphrases to optimize for memorability which is in line with earlier work~\cite{bonneau_linguistic_2012}. %
Second, we  show %
while system-generated passphrases today optimize primarily for security, users prefer passphrases with proper syntax over a random set of words due to memorability. Finally, our work reveals that there is a trade-off between memorability and guessability, and it is possible to balance these two factors for creating more usable system-generated passphrases. Our high-level observation is likely to translate to other non-English languages and culture-sensitive passphrases which is a fertile avenue for future work. %

\begin{acks}
We thank the anonymous reviewers and our shepherd Hyoungshick Kim for their valuable feedback. We also thank Abir De, Mohit Gupta and Husne Atiya for their help and input with an earlier iteration of this work. %
This research was (partially) funded by the Federal Ministry of
Education and Research (BMBF), Germany under the project LeibnizKILabor
with grant No.\,01DD20003 and by a Google India Faculty Research Award.
\end{acks}

\bibliographystyle{ACM-Reference-Format}
\bibliography{mascara}

\appendix
\section{Detecting user-generated passphrases}\label{sec:inusepassphrase}

\subsection{User passphrases}

To examine the class of user passphrases, we need to have a dataset compromising user passphrases. Unfortunately, unlike passwords, where data breaches are not uncommon and a lot of which have surfaced publicly~\cite{fouriq}, there is no public dataset of passphrases available. But we notice that several password leak databases contain long passwords that could potentially be passphrases without a proper delimiter. For this, we devise a segmentation algorithm to identify passphrases from password leak datasets and construct the first user-chosen, in-use passphrase database.  We
describe in this section how we extracted the passphrases and released the code/dataset from this paper for further research on passphrases in 
\begin{center}
\small \color{blue}{\tt \url{https://github.com/Mainack/MASCARA-passphrase-code-data}}
\end{center}

\paragraph{Password leak dataset.} We use a compilation of prior data breaches that surfaced in 2018 by 4iQ security firm~\cite{fouriq}. The leaked dataset contains nearly 1.4~billion email-password pairs. Prior research on passwords has used this leak~\cite{pal2019beyond,li2019protocols}. We use this dataset to extract potential in-use passphrases.

To find passphrases in this dataset, we only consider passwords that are longer than 20 characters, or roughly 4 words (given that the average length of an English word is 4.8~characters~\cite{Norvig}). We found 5.7~million unique passwords that are longer than 20 characters. These passwords were used by 5.9~million users (identified by email addresses in the dataset) \footnote{We combined all the passwords (of any length) belonging to the same email. We further combined the emails (and the corresponding passwords) if the two emails share the same username (the part before the `@' symbol) and if they have a common password. We ignored the users with more than 1,000 passwords, as they are unlikely to be real user accounts. Such preprocessing was also done in~\cite{pal2019beyond}.} 6.2~million times in total.

\paragraph{Segmenting passwords.}\label{sec:segmenting-pws}
As these passwords are selected from a compilation of password leaks, many of them are potentially hash values ~\cite{pal2019beyond}. We remove these hash values using a heuristic-based identification algorithm. We check if the password only contains hexadecimal characters and if so, we flag them as hash values and remove them. This removed 1.9~million unique passwords. We also find many of the 20-or-more character passwords look like email addresses or have some parsing errors. We removed such 1.5~million passwords that contain an `@' symbol with a prefix and has `.' in its suffix. We
also removed 0.4~million passwords that had less than nine English letters in them, as that is the minimum number of characters necessary to form a three-word passphrase, with each word at least three characters long (See details below). This left us with 1.8~million passwords that we then test using our passphrase segmentation algorithm.
We used a standard NLP task of doing word segmentation of noisy text for segmenting passwords and creating passphrases. Initially we tried segmenting using SymSpell library~\cite{garbe2019symspell} which is parameterized by a unigram distribution $V$ (created using popular word lists~\cite{NLPch02, geo, moby}). Given a password $\pp$, it can segment to create a sequence of words $(\w_1,\w_2,\ldots,\w_\pwlen)$, such that $\prod_{i=1}^{\pwlen} \Pr[\w_i]$ is maximized. SymSpell also tolerates a specified amount of variations of the words in $V$. However, this segmentation often failed to identify proper nouns like city names or first names which are often part of a password. Thus, we devised a hybrid approach. First, we implemented a greedy strategy to find known words in a password where we searched for words from exhaustive lists of known words including names of countries and popular cities~\cite{geo}, and common first names in the US~\cite{moby}. We only consider a password as passphrase if it has at least three valid English words, each of length greater than or equal to 3. If this greedy approach fails to segment a given passphrase (likely signifying variations of words), we used SymSpell library for segmentation.

\par
\begin{figure}[t]
  \centering
  \scriptsize
\begin{tabular}{p{0.7in}l}
  \toprule
  Type & Examples \\\midrule
  \multirow{-2}{0.7in}{Popular passphrases} & \begin{tabular}[c]{@{}l@{}}
                            \texttt{bullet-for-my-valentine}\\
                            \texttt{sponge-bob-square-pants}\\
                            \texttt{get-there-very-fast-indeed}
                            \end{tabular}    \\\cmidrule{2-2}
  \multirow{-2}{0.7in}{Unpopular passphrases} & \begin{tabular}[c]{@{}l@{}}\texttt{friendly-neighborhood-pickle} \\
                    \texttt{eddie-the-penguin-stick} \\
                    \texttt{super-looper-evil-ben}\end{tabular}    \\\midrule\midrule
  {Ineligible} & \begin{tabular}[c]{@{}l@{}}\texttt{speedtriple123456789} \\ \texttt{21101975-invalidlogin}\\ \texttt{newjob2thomapink\_socks08}\end{tabular} \\ \midrule
  \multirow{-1}{*}{Common names} & \begin{tabular}[c]{@{}l@{}}\texttt{KatherineCarrasquillo} \\ \texttt{zuleimahernandez1230} \\ \texttt{dobrovolskayatatiana}\end{tabular}             \\ \midrule
  {Non-phrasal} & \begin{tabular}[c]{@{}l@{}}\texttt{ltdjxrfgtctybz27102003} \\ \texttt{oilgurtalococsecnarf} \\ \texttt{903kingdalonsbfreitag}\end{tabular}           \\ \midrule
\end{tabular}
\caption{Figure shows different types of passwords that we analyzed. In the top row, we show the most common passphrases, as well as random samples that we were able to extract from passwords, while in the bottom three rows, we show the passwords that we do not consider as passphrase, and the categories they fall into.
}
\label{fig:ex-passphrases}
\end{figure}

\paragraph{Resulting passphrases dataset.}  Using our segmentation approach, we found~~$\nppws$ passphrases from our segmentation algorithm. We show the top three most used passphrases as well as three randomly sampled passphrases from the ones used by only one user in the top row of~\figref{fig:ex-passphrases}.

We also checked the passwords which were not considered as passphrases by our algorithm to gauge the false negative rate.  We random sampled hundred such passwords from 1.8~million passwords that are discarded by our algorithm and manully analyzed them to find three key types in this set: (a) \emph{Ineligible} (passwords having less than 3 words), (b) \emph{Common names}
(passwords that are common names, and our segmentation algorithm failed to segment them meaningfully), and (c) \emph{Non-phrasal} (passphrases that are just a mix of letters, digits, and symbols that do not segment into any meaningful sequence of words.) We show samples of these three types of passwords at the bottom three groups of rows in \figref{fig:ex-passphrases}. 

\subsection{Ecological validity of our dataset}\label{sec:ecolofical_passphrase}

\changed{As mentioned earlier, the passphrases we identified are a subset of a broader password breach dataset that is widely used in prior works and is known to contain real email-password pairs~\cite{fouriq}. According to the work by Bonneau et. al.~\cite{bonneau_linguistic_2012}, passphrases in their Amazon PayPhrase dataset (not publicly available) have linguistic properties similar to natural language. We make the same observation in our dataset too using a GPT LM-scorer, which shows user-generated passphrases that we extract are similar to natural text (\figref{fig:lm-cdf}) and some examples are given in \figref{fig:ex-passphrases}. %
We verified that the frequency distribution of the passphrase dataset mirrors that of the original breached password dataset, hinting at ecological validity of our passphrase dataset--three most frequent passphrases are used by 252 (0.26\%), 115 (0.12\%), and 102 (0.10\%) users.}

\section{Linguistic properties of in-use passphrases}\label{sec:perplexity}

\begin{figure}[t]

  \includegraphics[width=6.5cm]{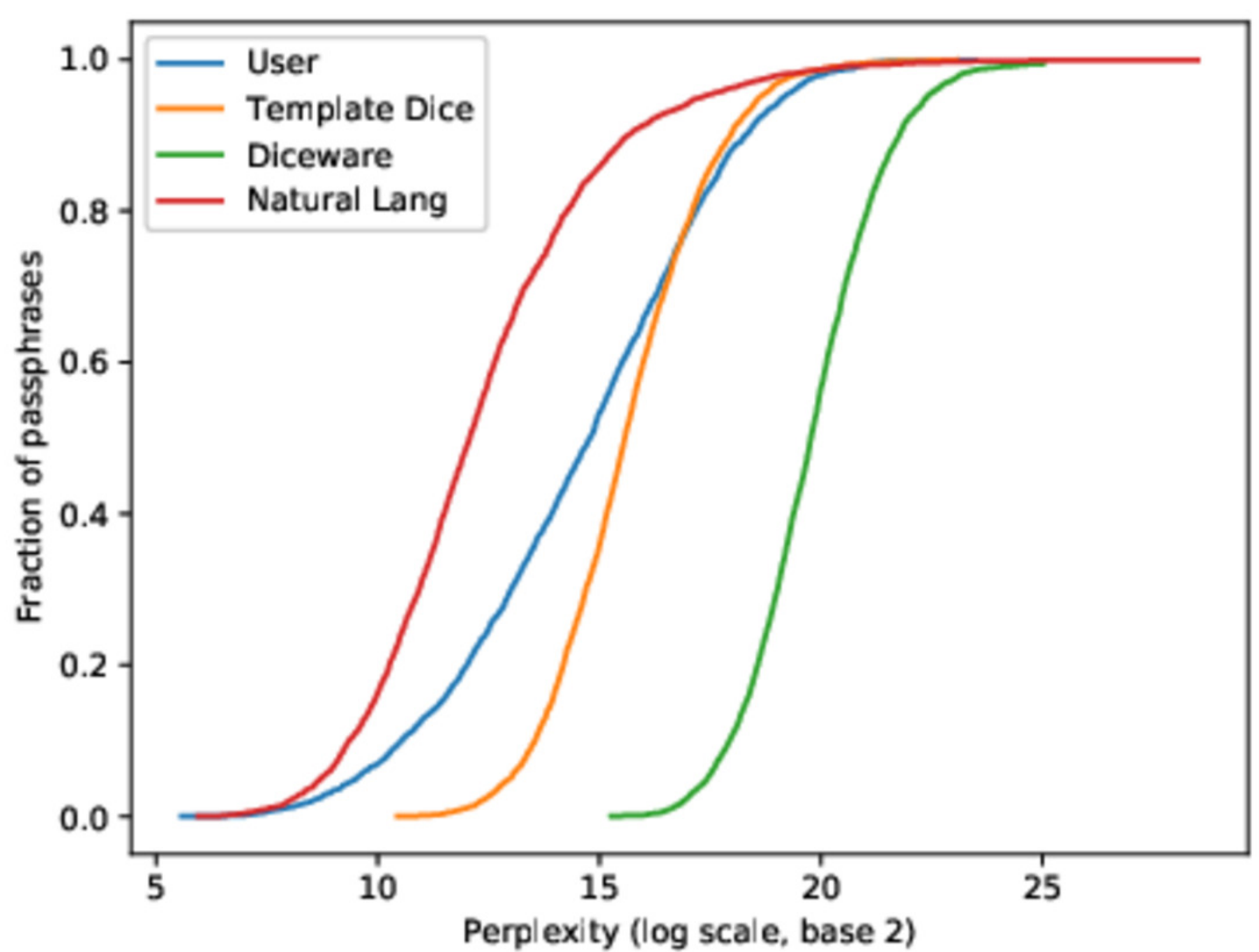}
  \caption{CDF of logarithm of perplexity to the base 2 for \dice, \mmap and \userpp passphrases, along with natural language phrases for baseline. 
  }
  \label{fig:lm-cdf}
\end{figure}

We leveraged a natural language model GPT-2~\cite{radford2019language} to check similarity of various system-generated passphrases as well as user passphrases to the natural language~\cite{radford2019language}. GPT-2 is trained to predict the next word given a sequence of words and widely used today~\cite{exgpt2}. We sampled 3000 passphrases of similar length distribution from each system (dataset for \userpp) and compute their perplexity using GPT-2. Lower perplexity implies similarity to natural language. The distribution of the perplexity score~\cite{brown-1992-perplexity} for both the system  (\dice and \mmap) and \userpp passphrases are shown in~\figref{fig:lm-cdf}. \userpp passphrases have a very low perplexity value which is similar to the perplexity of a natural language corpus, uncovering a key reason for their memorability. On the other hand, the passphrases from \mmap claim a close second in their resemblance to natural language, almost comparable to user passphrases, which indicates a significant improvement over the passphrases generated by \dice. %

\section{Shortcoming of \mmap}\label{sec:mmap-shortcoming}

\begin{figure}[t]
  \includegraphics[width=6.5cm]{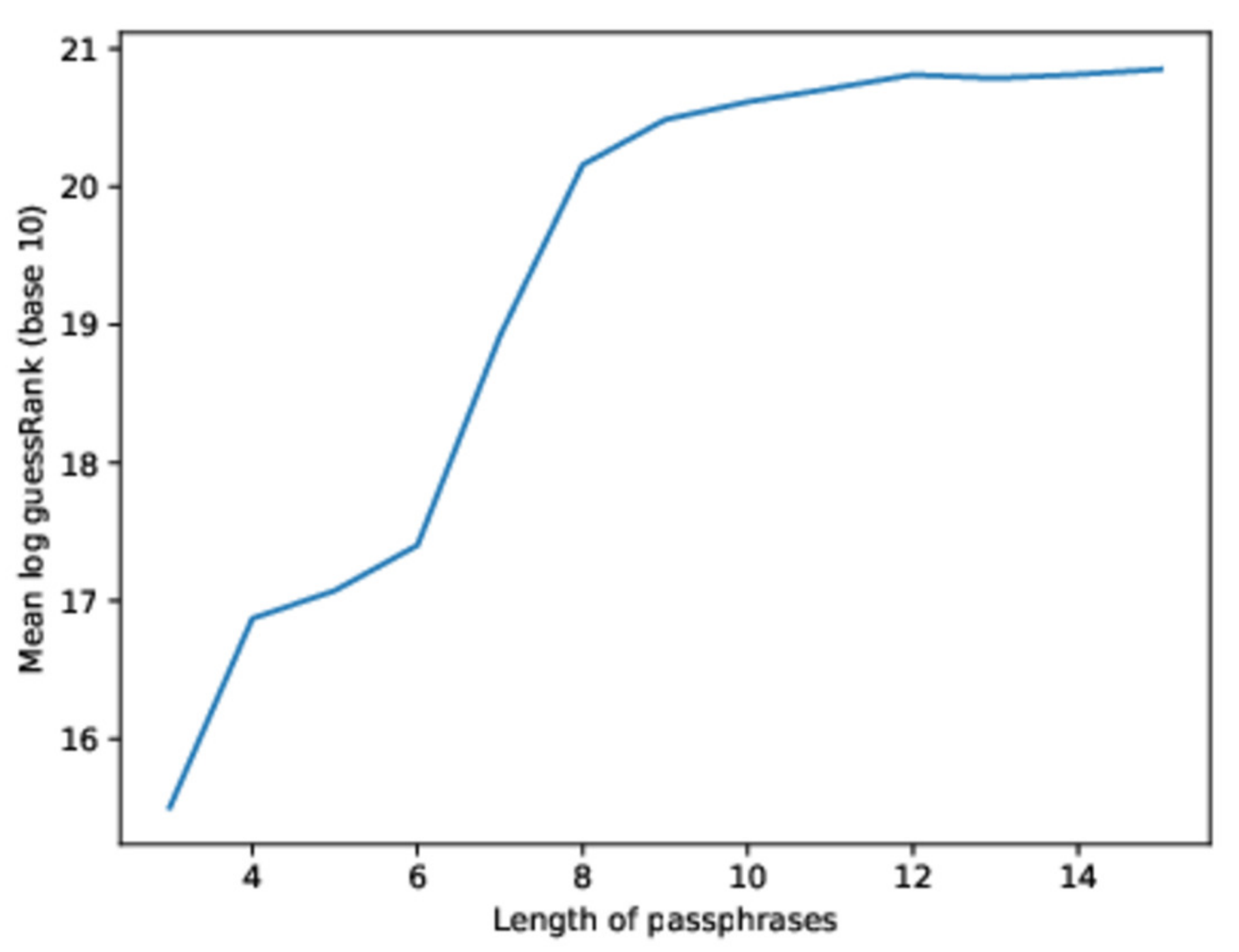}

  \caption{Variation of mean log$_{10}$ guessrank across length for the \mmap passphrases. We observe that guessranks reached a plateau after length eight. %
  }
  \label{fig:mmap-length}
\end{figure}

We show the guessrank of an adversary who does brute force guess of the syntax rules and then wordlists on \mmap passphrases in \figref{fig:mmap-length}. We note that the guessranks of these passphrases gets saturated around length 8---guessrank of 8-word passphrase is nearly the same as that as of 13-word. The potential reason is the constraint imposed by the underlying hardcoded and extremely limited syntax rule patterns of \mmap

\section{Trade-off for \system{} parameters}\label{sec:thresholdtradeoff} 

\noindent \textbf{Constraint thresholds.} \label{ssec:constraints} We introduce $\thetaone$ and $\thetatwo$ as thresholds to be satisfied by each of the probable words from the support (at every step of generation). Words that fail in any of the two constraints are removed from the support, and then we choose the next word weighted by their conditional bigram probabilities.

The goal of these empirical thresholds is to impose constraints on a relaxed upper bound estimate for the CER score, $\score$, and a relaxed upper bound estimate for the resultant bigram probability, $\lprobbi(w_{i-1},w_{i})$. They also ensure that generated passphrases can still retain syntactic structure and flow. 

Since our CER estimate is a linear fit, and the log probability of a phrase is sum of the log probabilities of its constituent unigrams or bigrams, we estimated the score obtained at every step as a greedy approach to obtaining a sub-optimal solution, rather than having to traverse through every path from the $\start$ token exhaustively. This approach is beneficial for using a different corpus or controlling the weight of the factors influencing the CER estimate.

\textbf{Equation coefficients: }$ \{\alpha_{i}\}_{i=1}^3$ are the coefficients of the individual parameters we fit with the regression model trained over the \wiki dataset as the universal corpus (\secref{memopt}). These essentially determine the weight of each factor in the estimation of a phrase's memorability. The upper bound of the log bigram probability, $\thetatwo$, is set to a suitable fraction of the minimum log bigram probability found in the corpus. Similarly, the threshold for  intermediate CER score ($\score$), $\thetaone$, is set as fraction of maximum CER.

\section{Example of passphrases from different algorithms}\label{sec:passsample}

\begin{figure}[t]
  \centering
  \scriptsize
  \begin{tabular}[t]{lrp{2.3in}}
    \toprule
    Type & &  Samples \\\midrule
    \multirow{5}{*}{\dice}
         & 1) & \texttt{dreamscape manchuria dervish verbally} \\
         & 2) & \texttt{clay reactive smasher authentic chrome hamster} \\
         & 3)&  \texttt{spindle chemicals griminess waviness vintage stammer agenda sulphate} \\
    \midrule
    \multirow{3}{*}{\userpp}
         & 1) & \texttt{mind the fold law you should} \\
         & 2) & \texttt{just another happy ending} \\
         & 3) & \texttt{dont cry over spilt milk} \\
    \midrule
    \multirow{4}{*}{\markov}
         & 1) & \texttt{leopold arranged for some users include the war} \\
         & 2) & \texttt{during ultraviolet signals beamed}\\
         & 3) & \texttt{their home delivery and morgan suggested that more} \\
    \midrule
    \multirow{5}{*}{\mmap}
         & 1) & \texttt{when does a bellboy spike an elect but not a sidebar} \\
         & 2) & \texttt{why does Suzy grumble a redeemer}\\
         & 3) & \texttt{how does my overdone one push those violinists after their sounding} \\
    \midrule
    \multirow{4}{*}{\ours}
         & 1) & \texttt{edge bands influenced how far north south} \\
         & 2) & \texttt{stalin however was offered exclusive control those four} \\
         & 3) & \texttt{graham and republic records} \\
    \midrule
    \multirow{4}{*}{GWC}
         & 1) & \texttt{rub revisions lilo clerk apple beting} \\
         & 2) & \texttt{helios hounds binary canonized lady overflight} \\
         & 3) & \texttt{pressures broth billable playgirl raita dekko} \\
    \bottomrule
  \end{tabular}
  \caption[Passphrase samples]{Three randomly sampled passphrases from each
    group of passphrases we consider for evaluation.}
  \label{fig:samples}
\end{figure}

\figref{fig:samples} presents a set of randomly chose passphrases generated by each algorithm and the ones generated by users.

\section{User Study instrument}
\label{sec:userstudy}

\subsection{Part 1}\label{ssec:part1}
\begin{footnotesize}

\textbf{Informed consent}

\noindent First we show the informed consent to the users. Users will see the questions below only if they indicated that they understood the requirements and agree to participate.

\begin{itemize}
\item Please enter your prolific id \_\_\_\_\_\_

\item When you login to your online accounts you often need a credential, i.e., your email/username and a secret. 

Two possible ways is creating the secret is: choosing a password (a set of characters) or choosing a passphrase (a set of words). Example of a password is "!Passw0rd!" and example of a passphrase is “correct horse battery staple”. 

You can also use a password or a passphrase as a master secret for accessing your password manager (a software that stores and manages all of your login credentials across multiple online accounts). 

Do you use passphrases as a secret for logging-in to any of your online accounts?
$\ocircle$ Yes $\ocircle$ No

\textcolor{gray}{if YES to use passphrases}
\item  Please briefly explain why do you use passphrases for these account(s) instead of passwords? (1-2 sentences) \hrulefill

\textcolor{gray}{if YES to use passphrases}

\item  How did you generate your passphrase(s)? Choose all that apply.

\begin{enumerate}
    \item [\ensuremath{\circ}]  Used an online tool [also write names, if you remember]: \hrulefill
    \item [\ensuremath{\circ}]  Self-generated - using a quote from a poem, movie, or book
    \item [\ensuremath{\circ}]  Self-generated - using a combination of random words
    \item [\ensuremath{\circ}]  Other: \hrulefill
\end{enumerate}

\textcolor{gray}{if NO to use passphrases}

\item Briefly explain why do you NOT use passphrases for these account(s) (and use passwords)? (1-2 sentences)

\textcolor{gray}{if NO to use passphrases}

\item If you have to use a passphrase as your master credential for a password manager, how would you generate it? Choose all that apply. 
\begin{enumerate}
    \item [\ensuremath{\circ}] Use an online tool [also write names, if you remember]: \hrulefill
    \item [\ensuremath{\circ}] Self-generated - will use a quote from a poem, movie, or book
    \item [\ensuremath{\circ}] Self-generated - will use a combination of random words
    \item [\ensuremath{\circ}] Other: \hrulefill
\end{enumerate}

\item Which of these crawls?
$\square$ Dog $\square$ Cat $\square$ Snake $\square$ Kite

\item  For the questions below please choose the option which applies most for you
\begin{itemize}
    \item Do you write down your login credentials to remember them?\\
    Never \ensuremath{\circ} \ensuremath{\circ} \ensuremath{\circ} \ensuremath{\circ} \ensuremath{\circ} Always
    \item How often do you use the same  login credential for multiple websites?\\
    Never \ensuremath{\circ} \ensuremath{\circ} \ensuremath{\circ} \ensuremath{\circ} \ensuremath{\circ} Always
\end{itemize}

\noindent \textbf{Passphrase choice}

\item In this section, we will show you a few passphrases and ask you to choose one.  A good passphrase should be long so that others cannot guess it, but also be easy to remember so that you can enter the passphrase with minimum errors.

We are showing three passphrases and how secure they are in terms of the time it might take to guess the passphrase by an attacker. Please choose one passphrase which you prefer as your master login credential (e.g., for your password manager). %

Please do not write down the chosen passphrase.%
Try to remember it to the best extent possible. %

Please DO NOT COPY/PASTE passphrases in this study. Such actions will be detected and your task could be invalidated.

    [Passphrase 1] [Passphrase 2] [Passphrase 3]

\textcolor{gray}{Repeat question below 5 times for practicing}
\item For practicing, please enter your chosen passphrase: [CHOSEN PASSPHRASE] (4 more practices remaining)

\noindent \textbf{Post Passphrase choice questions}

\item Why did you choose this particular passphrase among the ones shown? (1 - 2 sentences): \hrulefill

\item Please select options from below which have positively affected your memorability of the chosen passphrase
    \ensuremath{\circ} Contains frequently used words.
    \ensuremath{\circ} Grammatically correct
    \ensuremath{\circ} Fewer words to remember as it contains common words
    \ensuremath{\circ} Flows like an English phrase
    \ensuremath{\circ} Other: \hrulefill

\noindent \textbf{Demographics}

\item Which age group do you belong to?
          \ensuremath{\circ} 18-30 
     \ensuremath{\circ} 31-45
     \ensuremath{\circ} 46-60
     \ensuremath{\circ} 60+

\item Which gender do you identify yourself most with?
    \ensuremath{\circ} Male 
    \ensuremath{\circ} Female
    \ensuremath{\circ} Non-Binary / Third Gender
    \ensuremath{\circ} Prefer not to say

\item What is the highest degree or level of school you have completed?
    \ensuremath{\circ} Some high school 
    \ensuremath{\circ} High school  
    \ensuremath{\circ} Some college  
    \ensuremath{\circ} Trade, technical, or vocational training 
    \ensuremath{\circ} Associate’s degree 
    \ensuremath{\circ} Bachelor’s degree
    \ensuremath{\circ} Master’s degree 
    \ensuremath{\circ} Professional degree 
    \ensuremath{\circ} Doctorate 
    \ensuremath{\circ} Prefer not to say 

\item Which of the following best describes your educational background or job field?
    \ensuremath{\circ} I have an education in, or work in, the field of computer science, engineering, or IT.
    \ensuremath{\circ} I do not have an education in, or work in, the field of computer science, engineering, or IT. 
    \ensuremath{\circ} Prefer not to say.

\end{itemize}
\end{footnotesize}

\subsection{Part 2} \label{ssec:part2}
\begin{footnotesize}

Welcome to a brief follow-up of the earlier study on passphrases that you completed. Recall that our study is on understanding utility of passphrases as login credentials for online accounts. 

Your primary task in this final part of this study is to just re-enter your chosen passphrase as you remember in this survey and answer a few questions regarding your current perception about passphrases. 
This part will take around 3 to 5 minutes of your time. You’ll be compensated \$1.00 USD for this part. Thank you for helping in our research to improve the privacy and security of users by understanding usage of passphrases.
\noindent\textbf{Questions}

\begin{itemize}

\item To start the survey please enter your Prolific ID \hfill 

\noindent \textbf{Check recall after 2 days}

\textcolor{gray}{Repeat the question below at most five times}

\item Please enter the passphrase you chose in Part 1 of this study below. (5 tries remaining). \_\_\_\_\_\_\_\_\_\_\_\_

\item What is your preferred length for a passphrase you would want to use (the number of words)?

\ensuremath{\circ} 7 or less words \ensuremath{\circ} 8-12 words \ensuremath{\circ} $>$12 words

\item Please briefly explain your choice of preferred length for passphrases (1-2 sentences): \hrulefill

\item If you have to use a passphrase, then while choosing the passphrase, would you prefer to prioritize security or memorability? 
\\Prioritize only Memorability \ensuremath{\circ} \ensuremath{\circ} \ensuremath{\circ} \ensuremath{\circ} \ensuremath{\circ} Prioritize only Security

\item Currently, do you feel changing your chosen passphrase slightly (e.g., a few characters) would have made it more memorable for you without making it easy to guess for others?
\begin{enumerate}
    \item [\ensuremath{\circ}] No, I don’t want to change the passphrase.
    \item [\ensuremath{\circ}] Yes, I want to change my chosen passphrase to \hrulefill
\end{enumerate}

\item In brief, why do you feel your modification to the chosen passphrase will make the passphrase more memorable without making it easy to guess for others? (1-3 sentences) \hrulefill

\item After participating in this study, how likely are you to use passphrases as your login credential for some online accounts instead of passwords?\\
Not Likely At All \ensuremath{\circ} \ensuremath{\circ} \ensuremath{\circ} \ensuremath{\circ} \ensuremath{\circ} Very Likely

\item Please briefly explain your response (1-3 sentences) \hrulefill

\end{itemize}

\end{footnotesize}

\section{Comparing \system{} with human-in-the-loop passphrase generation}\label{sec:gwc}

\changed{We further 
tested the utility of \ours against passphrase generation systems which puts human in the loop (and presumably incur higher cognitive and time cost for users in exchange for more user control). Specifically, we compared the memorability of \ours generated passphrases with passphrases generated via guided word choice (GWC) method %
~\cite{blanchard-2018-gwc}.  GWC asks users to choose words for their passphrase from a list of random words shown to them.    
On our request, Blanchard et al. graciously shared the set of passphrases used in their experiment~\cite{blanchard-2018-gwc} (Implementation of GWC is unavailable). We show a sample of GWC passphrases in~\figref{fig:samples}. Thus, in this experiment we repeated part 1 of our original study (since Blanchard et al. also did not ask participants for a followup task~\cite{blanchard-2018-gwc}). A total of 37 users (recruited via mailing lists) were randomly shown either \ours generated passphrase or GWC-generated passphrases (from original paper). Then these users typed the chosen passphrases via a login screen after answering a few distracting questions. (18 users were shown GWC passphrases and 19 were shown \ours{} passphrases). A comparable 72.2\% users recalled GWC passphrases correctly in one try and 73.7\% users recalled \ours passphrases in one try. However, interestingly, the average edit distance for GWC passphrases was 2.72 which is higher than the average edit distance of \ours passphrases (1.46). Thus, although GWC provides users more control while choosing random words, in our experiments \ours-passphrases are still slightly more memorable (and less error prone) than GWC-generated passphrases.}

\end{document}